\documentclass[
reprint,
superscriptaddress,
amsmath,amssymb,
aps,
prb,
floatfix,
]{revtex4-2}
\usepackage{comment}
\usepackage{bm}
\usepackage{tikz}
\usepackage{tikz-feynman}
\usepackage{graphicx}
\usepackage{booktabs}
\usepackage{bm}
\usepackage{braket}
\usepackage[colorlinks]{hyperref}
\usepackage{color}
\usepackage[usenames,dvipsnames]{xcolor}
\usepackage[export]{adjustbox}
\usepackage{siunitx}
\usepackage{fouriernc}
\usepackage{float}
\usepackage{placeins}
\usepackage{pgfplots}
\pgfplotsset{compat=1.18}
\usepackage{xcolor}
\usepackage{pgfplots}
\pgfplotsset{compat=newest}

\emergencystretch=\maxdimen
\newcommand{\be}{\begin{equation}}
\newcommand{\ee}{\end{equation}}

\begin{document}
	
	\small
	
	\title{Bipartite entanglement under frequency comb pumping in parametric Josephson circuits}
    
	\author{Mikael Vartiainen} \email{mikael.vartiainen@aalto.fi}
	\affiliation{Low Temperature Laboratory, Department of Applied Physics, Aalto University, P.O. Box 15100, FI-00076 Espoo, Finland}
    \thanks{Corresponding author: Mikael Vartiainen}

    \affiliation{QTF Centre of Excellence, Department of Applied Physics, Aalto University, P.O. Box 15100, FI-00076 Aalto, Finland}
 
	\author{Ilari Lilja}
	\affiliation{Low Temperature Laboratory, Department of Applied Physics, Aalto University, P.O. Box 15100, FI-00076 Espoo, Finland}
	
	\affiliation{QTF Centre of Excellence, Department of Applied Physics, Aalto University, P.O. Box 15100, FI-00076 Aalto, Finland}
 
	\author{Ekaterina Mukhanova}
	\affiliation{Low Temperature Laboratory, Department of Applied Physics, Aalto University, P.O. Box 15100, FI-00076 Espoo, Finland}
	
	\affiliation{QTF Centre of Excellence, Department of Applied Physics, Aalto University, P.O. Box 15100, FI-00076 Aalto, Finland}%

    \author{Kirill Petrovnin}
	\affiliation{Low Temperature Laboratory, Department of Applied Physics, Aalto University, P.O. Box 15100, FI-00076 Espoo, Finland}
	
	\affiliation{QTF Centre of Excellence, Department of Applied Physics, Aalto University, P.O. Box 15100, FI-00076 Aalto, Finland}%
    
    \author{Gheorghe Sorin Paraoanu} \email{sorin.paraoanu@aalto.fi}
	\affiliation{Low Temperature Laboratory, Department of Applied Physics, Aalto University, P.O. Box 15100, FI-00076 Espoo, Finland}

    \affiliation{QTF Centre of Excellence, Department of Applied Physics, Aalto University, P.O. Box 15100, FI-00076 Aalto, Finland}

	\author{Pertti Hakonen}  \email{pertti.hakonen@aalto.fi}
    \affiliation{Low Temperature Laboratory, Department of Applied Physics, Aalto University, P.O. Box 15100, FI-00076 Espoo, Finland}
    
    \affiliation{QTF Centre of Excellence, Department of Applied Physics, Aalto University, P.O. Box 15100, FI-00076 Aalto, Finland}
	\date{\today}%
	
	\begin{abstract}		
The creation of high-quality cluster states in superconducting microwave circuits is a relevant ingredient in continuous-variable quantum computing. Although large-scale cluster states have been established in optical systems, dissipation prevents their direct applicability to the microwave realm. Recent improvements in superconducting parametric circuits, in particular Josephson parametric amplifiers (JPA) and traveling wave parametric amplifiers (TWPA), have permitted substantial progress in producing entangled states using microwave photons. In this paper, we examine experimentally and theoretically the effects of numerous parametric pump tones on the degree of two-mode squeezing in a quantum circuit and apply it to the JPA. We find that additional pumps diminish the initial two-mode correlations achieved with a single pump by redistributing it among a larger network of modes and by introducing entanglement with additional idler frequencies. Taking into account the actual heterodyne measurement conditions, the experimental results are consistent with theoretical expectations.

	\end{abstract}
	
	\maketitle

\section{Introduction} \label{Intro}

The generation of continuous variable (CV) cluster states in superconducting microwave circuits attracts growing interest as a new alternative route towards universal quantum computing (QC) \cite{Zhang2006,Menicucci2006,Gu2009}. The entanglement between separate photonic modes provides a resource for advanced information processing relying on quantum mechanical concepts for one-way computation \cite{raussendorfbriegel, RevModPhys.81.865}. In this scheme, quantum measurements are employed for information processing instead of qubit-based gates. For advanced one-way quantum computing, highly entangled states are required for this scheme to challenge other approaches for QC. Highly entangled states have been achieved in quantum optics, where the generation of cluster states can be realized utilizing optical parametric amplifiers or oscillators. Parametric oscillators can be employed as squeezed light sources that can be combined using beam splitters and delay lines to produce a two-dimensional cluster state up to 30000 nodes \cite{Larsen2019}.

To perform one-way quantum computation, a complex cluster state with genuine multipartite entanglement is needed, and to make such a state in a CV platform, such as the JPA or JTWPA, one needs multiple parametric pumps. This can be achieved by either pulsating or continuous pumps, which modulate the coupling to the external line \cite{bruschi2016towards}. However, applications that rely on optical delay lines to generate cluster states do not directly translate to the microwave domain, due to attenuation and limited coherence times. 

Graph representations of quantum states have been used in the context of graph and cluster states \cite{cluster, multipartygraph, graphresource, square-ladder-cluster, alocco2025programmablemicrowaveclusterstates}.
For closed systems, a graph representation of a quadratic interaction Hamiltonian contains all the relevant information about how correlations are distributed among the system and allows a clear visual illustration of all interactions between different modes. Many global and graph-based entanglement measures have been calculated from the properties of such graphs, such as the nullifier-based criteria or H-graph formalism \cite{detectingmultiparty, menicucci2011graphical, multipartygraph, Menicucci2006, graphresource}, but these lie beyond to scope of this work. However, in reality, states are rarely pure, and entanglement easily becomes weakened by noise from the environment. 
In some cases, different measurement schemes can strengthen or weaken the correlations between different parts of your system \cite{serafinibook, pedagocicalmonitoring}. Graph and cluster states have also been used in the construction of quantum error correction codes \cite{Schlingemann_2001, Anders_2006}.

A key feature of multipump operation is that each additional pump introduces new idler modes, transforming an initially isolated two-mode squeezing interaction into a correlated multimode network \cite{square-ladder-cluster}. In such systems, entanglement is no longer confined to a single mode pair but becomes distributed among many modes. From general considerations of correlation sharing in Gaussian systems, one therefore expects that increasing the number of coupled modes redistributes the bipartite entanglement between any selected pair, thus reducing the individual bipartite entanglement \cite{detectingmultiparty, menicucci2011graphical, multipartygraph, monogamysource1, monogamysource2}. Quantifying this redistribution is essential for understanding the scalability of pump-engineered cluster-state architectures.

Owing to the large dissipation of microwave signals, the quantum optics methods are not directly applicable to the microwave domain, and other solutions are necessary. Recently, great progress has been made in superconducting quantum parametric circuits, both with Josephson parametric amplifiers (JPA) \cite{hernandezhaviland, havilandfreqcomb,  Yurke1988, Yamamoto2008, Eichler2011, Hatridge2011, Lahteenmaki2012, Mutus2013, Lahteenmaki2014, Mutus2014, Zhou2014, Roy2015, Jebari2018, Elo2019} and with traveling wave parametric amplifiers (TWPA)  \cite{Macklin2015,White2015,Zorin2016,Perelshtein2022,Esposito2022twpa, ranadive2024traveling,Qiu2023,livreri2024josephson,Nilsson2024TWPA,gaydamachenko2025rf}. Two-mode squeezing (TMS) up to -12 dB and -9.5 dB has been demonstrated in JPA  circuits \cite{Eichler2014} and TWPAs \cite{qiu2023broadband}, respectively. In TWPAs, two-mode squeezing with logarithmic negativity of $E_N\sim 0.8$ has been demonstrated over 4 GHz bandwidth \cite{Perelshtein2022}, a full octave in the microwave domain. Furthermore, genuine multipartite entanglement has been reached in JPA systems using more than one single pump \cite{square-ladder-cluster, alocco2025programmablemicrowaveclusterstates,  PetrovninSPDC}. The starting point of all these works is simple and easily accessible, that is, the generation of entanglement from the vacuum state of a quantum field \cite{SUMMERS1985257}.



In this work, we theoretically investigate how bipartite entanglement evolves as parametric pumping progressively enlarges the interaction network and apply the solution to our experiments done on the JPA. Pairwise entanglement is the basic building block of a cluster state and we use it as a probe of how correlations spread across a parametrically generated multimode cluster state. By adding up to fifteen pump tones and changing pumping strength, we systematically increase the connectivity of the system and measure the logarithmic negativity of a chosen pair of modes. We further compare symmetric and asymmetric pumping configurations, which allows us to examine how distinct coupling topologies and the sheer number of idler couplings influence the redistribution of entanglement. In the symmetric configuration, new pumps create correlations among a fixed mode set, resulting in a large number of beam splitter (BS) correlations. A BS correlation arises between two modes if both are TMS correlated to another same mode. In the asymmetric case, as new pumps introduce new idlers via TMS and do not operate on the fixed mode set, the number of idler modes is much larger, which leads to a greater loss of information, even though the number of BS correlations is much smaller. However, the trade-off with the symmetric pumping is that, due to the BS interactions, the system is much more phase-sensitive. Our goal is to understand the effects of these phenomena on bipartite entanglement inside a complicated network. 

Our measurements, supported by a Gaussian conditional-dynamics model including measurement backaction, show a rapid reduction of pairwise entanglement as additional modes participate in the interaction. We do not see significant difference between this reduction for either the symmetric or asymmetric cases, which allows us to compare the effect of BS correlations vs number of idler modes on the bipartite entanglement. Our calculations show that the relative phases of the pumps play a major role in the bipartite entanglement when there are lots of BS correlations.

The results demonstrate that multipumping redistributes entanglement across the network rather than generating pairwise squeezing, providing insight into the trade-off between multimode connectivity and usable bipartite resources in microwave cluster-state platforms. Our findings indicate that maintaining high entanglement between specific mode pairs becomes increasingly challenging as more pumps are introduced, as quantum information is redistributed across a larger network of modes. 

The remainder of this paper is organized as follows. Section~\ref{Experiment} introduces the experimental setup and analysis procedure. Section~\ref{OpenSyst} presents the theoretical framework for conditional and unconditional Gaussian dynamics, where we closely follow the approach of Ref. \onlinecite{serafinibook}. Section~\ref{GraphTheory} develops the graph-theoretical description of pump-induced couplings and their symmetry properties. Section~\ref{ResultsDisc} compares experiment and theory for up to fifteen pumps. Conclusions are given in Sec.~\ref{Conclusion}, with additional technical details provided in the Appendices.

\section{Experimental Techniques} \label{Experiment}

\begin{figure}[tb]
    \centering
    \includegraphics[width=\linewidth]{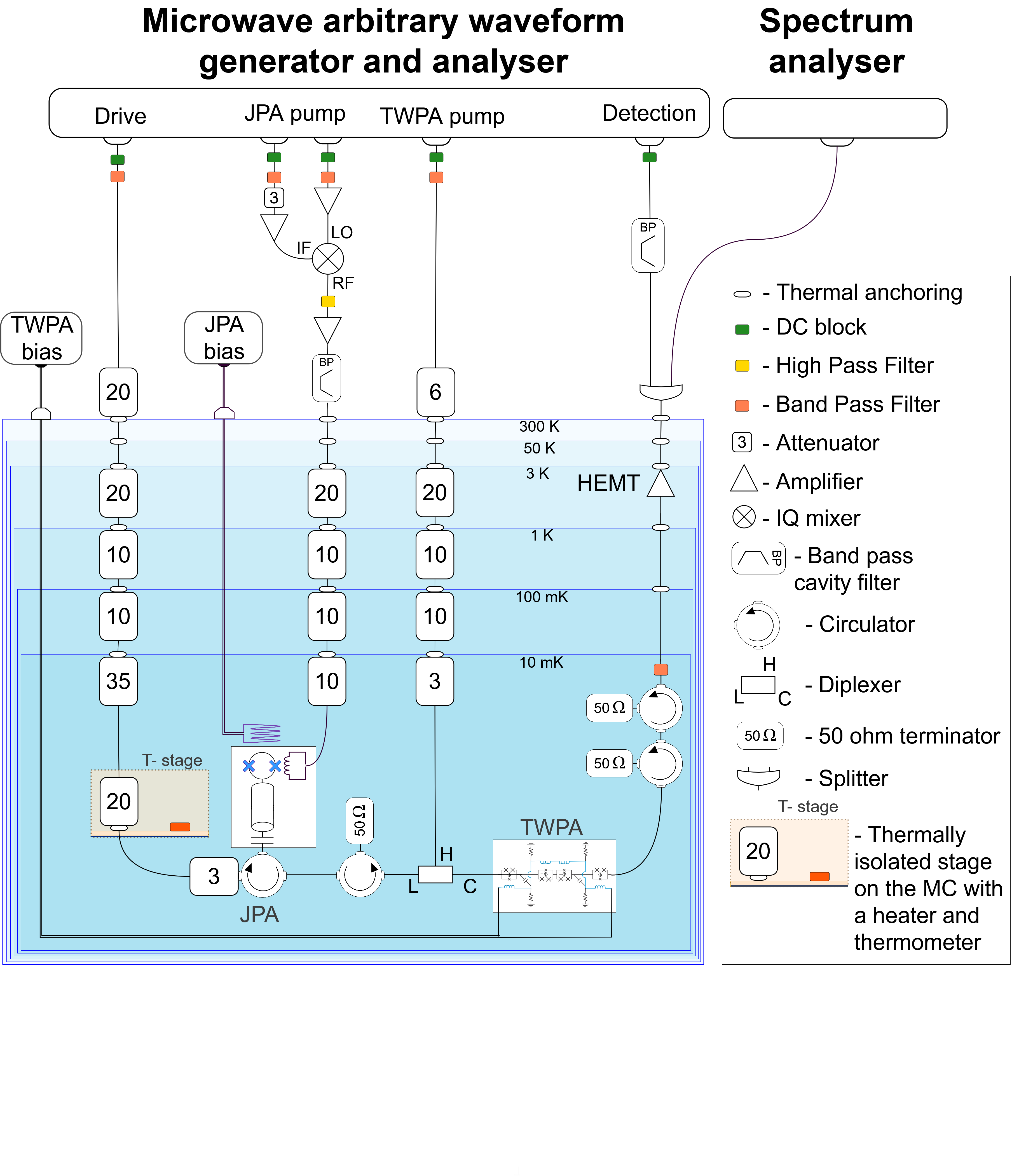}
    
    \caption{Schematic of the experimental microwave setup working around ~5.8 GHz frequency. The experiment is controlled by a GHz frequency lock-in signal analyzer, which also generates the required parametric pumps. The employed components are denoted in the frame on the right. The detection is performed using digital heterodyning \cite{PrestoIP}. For details of the operation, see text. 
    }
    \label{Fig:ExpSetup}
\end{figure}

The measurement setup is illustrated in Fig.~\ref{Fig:ExpSetup}. The source of entanglement in this work is the Josephson parametric amplifier (JPA) which was fabricated using a trilayer Nb process \cite{Gronberg2017}. The JPA is a quarter-wave cavity terminated by a SNAIL loop, used for frequency tuning over the working range of $6.1 - 5.5$ GHz \cite{PetrovninSPDC}. SNAIL consists of two large junctions with critical current $I_c = 14\mu{}A$ and $\alpha=0.29$ giving the smaller junction critical current ratio \cite{frattini2018optimizing}. While an external magnetic coil provided the DC flux bias for operating point tuning, an on-chip bias line with a mutual inductance of $\sim 0.5 $pH to the SQUID loop was used for RF driving at twice the cavity frequency -- 3 wave mixing process. 

The JPA is operated in the reflection mode using a circulator to separate incoming and outgoing signals. A TWPA based on SNAIL unit cells is used as a preamplifier \cite{Perelshtein2022}. A 20 dB attenuator in the input line (T-stage in Fig.~\ref{Fig:ExpSetup}) acts as a variable-temperature Johnson noise generator which is used for calibration of the gain and noise of the amplifier chain. The extra 3 dB attenuator after the thermally controlled plate improves the heat sinking in case the platform does not cool down to the base temperature. 
%
The pump drive to the TWPA is applied via a Mini-circuits ZDSS-7G10G-S+ diplexer. The backaction from TWPA on the sample is removed using a circulator operating at the band $4-12$ GHz. The noise temperature of the TWPA ($T_N \sim 0.7$ K) includes the $< 1.0$ dB losses between TWPA and JPA.

To guarantee phase coherence between all pump signals and down-converted modes, we used a Presto arbitrary waveform generator and analyzer for microwaves \cite{Presto}, which makes it also possible to repeat and average the same measurement several times under identical conditions even when using complicated pump and mode configurations. Our Presto unit generates signals up to 9 GHz by direct frequency synthesis utilizing upper Nyquist bands. Owing to power loss with the higher frequency bands, amplifiers were added to the signal line driving the JPA (see Fig. \ref{Fig:ExpSetup}). The frequency comb for parametric pump signals was produced using a triple-balanced mixer, with a cavity filter singling out the lower sideband spectrum. The relative phases of pump tones were left to their default values and are thus not controlled in the experiment. As such, we use random values for the phases in the simulation.

The internal cavity loss rate $\gamma$ and the coupling coefficient $\kappa$ of the cavity to external 50 $\Omega$ environment were determined by fitting the S21 measurement data. From the fits, we obtained $\kappa/2\pi\approx 6$ MHz and $\gamma=0.21\kappa$.  The JPA is flux-tuned to $0.41\Phi_0$, which corresponds to a resonance frequency of $f_r=5.808$ GHz. The resonance bandwidth corresponds to $\Delta f = 2 \kappa /2\pi = 12$ MHz which yields $Q \simeq 500$ for our cavity. As the JPA operates in the three-wave mixing mode, the half-pump frequency of the fundamental TMS pump is chosen to be at the center of the cavity resonance. In the symmetric and asymmetric configurations, the cavity frequency is not occupied and the first modes (-1,1) are set at [-0.05, 0.05] MHz detuning from the resonance frequency (see Fig. \ref{Fig:30 mode sym} for the indexing of the modes). This spacing means a full span of $2\times(0.05 + 21 \times 0.1)=4.3$ MHz for 44 equally spaced modes (see App. \ref{mode_posit} for placement of the modes within the JPA resonance curve). All experimental results were obtained at the base temperatures of a Bluefors LD400 cryostat ($T \sim 10$ mK). 

The frequency dependence of the external coupling $\kappa$ and internal losses $\gamma$ was characterized independently through calibration measurements. Within the frequency window relevant for the pumped modes, the damping rates vary smoothly, and no sharp impedance resonances were observed. These calibrated values are used as input parameters in the theoretical model. $\kappa$ and $\gamma$ are not frequency dependent. The offset from cavity resonance frequency affects the damping rate, though this behavior is well modeled by the effective Hamiltonian already. We also keep our measured modes around the cavity resonance frequency initially, though higher pump amplitudes result in some shift in cavity frequency.

\section{Quantum dynamics of covariance matrix} \label{OpenSyst}

We are interested in the quantum dynamics of the JPA, weakly coupled to a dissipative environment. The environment of the JPA is taken as Markovian because its noise is white (delta-correlated in time), and practically no reflection from the environment reaches the JPA at the frequencies of interest. In our theoretical treatment, we model the internal dissipation as losses before the measurement, described by the quantum limited attenuator channel \cite{serafinibook, gaussianquantumchannels}. 
Our approach is based on the diffusive dynamics of the Gaussian covariance matrix, which is monitored by heterodyne measurements \cite{serafinibook, wiseman2009quantum}. We employ a standard Hamiltonian description for Spontaneous Parametric Down-Conversion (SPDC) occurring inside parametric cavities. Using the full quadratic Hamiltonian, we derive a formal solution to the steady-state of the monitored covariance matrix, which embodies the entanglement of interest. 


Gaussian states are described by a quadratic Hamiltonian. A useful property of Gaussian states is that the states can be completely described by their first and second moments. The second moments contain information on the correlations between different parts of the system, and entanglement properties can be evaluated. Another useful property is that Gaussian noise channels preserve the "Gaussianity" of the states: if a Gaussian state is put through a Gaussian noise channel, the state is still completely determined by the first and second moments. 

\subsection{System description}

We consider a quantum system of $n$ bosonic modes coupled to a dissipative output channel with a coupling rate $\kappa$ \footnote{In typical microwave cavity setups, $2\kappa$ sets the bandwidth. However, here we assume that all pump interactions are equally effective, regardless of frequency separation.}
In the Born-Markov regime \cite{serafinibook, opensysytemsbook}, the environment is weakly coupled and unaffected by correlations induced by the pumps.

We model a mode at frequency \( f_m \) as a delta function \( \delta(f - f_m) \), extendable to finite-bandwidth modes \cite{LoudonBook}. Further, we define creation and annihilation operators for each frequency mode. The creation and annihilation operators form the basis:
\begin{equation}
    \tilde{a} = (\hat{a}^\dagger_{-1}, \hat{a}^\dagger_{1}, \hat{a}^\dagger_{-3}, ..., \hat{a}_{-1}, \hat{a}_{1}, \hat{a}_{-3}, ...)^T.
\end{equation}
We define quadratures as \( \hat{x}_k = (\hat{a}^\dagger_k + \hat{a}_k)/\sqrt{2} \) and \( \hat{p}_k = i(\hat{a}^\dagger_k - \hat{a}_k)/\sqrt{2} \), forming the vector:
\begin{equation}
    \tilde{r} = (\hat{x}_{-1}, \hat{p}_{-1}, \hat{x}_{1}, \hat{p}_{1}, \hat{x}_{-3}, \hat{p}_{-3}, \hat{x}_{3}, \hat{p}_{3}, ...)^T.
\end{equation}
Bath operators \( \tilde{r}_B \) are defined analogously. We take the bath to consist of \( n \) independent modes. The coupling Hamiltonian is:
\begin{equation}
    \hat{H}_{C} = i\sqrt{\kappa} \sum_k (\hat{a}_k \hat{b}^\dagger_k - \hat{a}^\dagger_k \hat{b}_k),
\end{equation}
with bath quadratures \( \hat{x}_{B,k} \), \( \hat{p}_{B,k} \).

Instantaneous-time commutators (white noise) are assumed:
\begin{equation}
    [\tilde{r}_B(t), \tilde{r}_B^T(t')] = i\boldsymbol{\Omega} \delta(t - t'),
\label{commutation}
\end{equation}
where \( \boldsymbol{\Omega} = \bigoplus_n \begin{pmatrix} 0 & 1 \\ -1 & 0 \end{pmatrix} \) is the symplectic form. To ensure canonical commutation relations, we define the infinitesimal bath operators:
\begin{equation}
    \delta \tilde{r}_B(t) = \int_t^{t+\delta t} \tilde{r}_B(z) dz,
\end{equation}
with the commutator:
\begin{equation}
    [\delta \tilde{r}_B(t), \delta \tilde{r}_B^T(t')] = i\boldsymbol{\Omega} \delta t.
\end{equation}
Taking the limit $\delta t \rightarrow dt$, we get \( \delta \tilde{r}_B(t) = \tilde{r}_B(t)dt \equiv \tilde{R}_B(t) \sqrt{dt} \), the canonical structure is restored. \( \tilde{R}_B \) are quantum Wiener processes; below, we relabel them as \( \tilde{r}_B \). Notice that now $\tilde{R}_B$ must have dimensions of $\left[1/\sqrt{\kappa}\right] = \left[\sqrt{t}\right]$.

We can justify the above assumptions by the following properties of our experimental setup. First, we treat the dissipation channels as Markovian and frequency independent within the experimentally relevant bandwidth. This approximation is valid because the mode spacing and parametric bandwidth are small compared to the inverse bath correlation time and to the scale over which the external coupling varies. Second, we neglect higher-order nonlinear mixing terms in the Hamiltonian beyond the leading parametric interactions generated by the applied pump tones. In the operating regime of our experiment, these higher-order terms are perturbatively suppressed and remain below the measurement resolution. Finally, strong pumping can induce mode hybridization and small resonance shifts via Kerr-type and higher-order effects. Residual impedance structure of the measurement chain leads to weak frequency-dependent variations of the damping rates; however, these variations are smooth over the frequency window considered and do not qualitatively modify the covariance dynamics.

\subsection{Equations of motion for the covariance matrix}

The full Hamiltonian of the combined system can be expressed compactly:
\begin{equation}
\hat{H} = \frac{1}{2} \tilde{r}^T H_S \tilde{r} + \tilde{d}^T \boldsymbol{\Omega} \tilde{r} + \tilde{r}^T C \tilde{r}_B,
\label{hamiltonian}
\end{equation}
with \( C = \sqrt{\kappa}\, \boldsymbol{\Omega}^T \).

The covariance matrix of a Gaussian state is defined as:
\begin{equation}
\boldsymbol{\sigma}_{ij} = 2 \langle (r_i - \langle r_i \rangle)(r_j - \langle r_j \rangle) \rangle.
\end{equation}

By considering the evolution under the full Hamiltonian in Eq. \ref{hamiltonian} over an infinitesimal time via a symplectic transformation on the covariance matrix $\mathbf{\sigma}$, one can derive the  \textit{unconditional} equation of motion \cite{serafinibook}:
\begin{equation}
\dot{\boldsymbol{\sigma}} = A\boldsymbol{\sigma} + \boldsymbol{\sigma} A^T + D,
\label{uncond}
\end{equation}
where the drift and diffusion matrices $A$ and $D$ are given by:
\begin{align}
A &= \boldsymbol{\Omega} H_S + \frac{1}{2} \boldsymbol{\Omega} C \boldsymbol{\Omega} C^T, \\
D &= \boldsymbol{\Omega} C \boldsymbol{\sigma}_B C^T \boldsymbol{\Omega}^T.
\end{align}

Following \cite{serafinibook}, this can be done by first assuming that the system and the environment are initially uncoupled $\mathbf{\sigma}=\mathbf{\sigma}_S\oplus\mathbf{\sigma}_B$. For simplicity, consider only the system-environment interaction, described by the matrix: 

\begin{equation}
    H_C = 
\begin{pmatrix}
    0 & C \\
    C^T & 0 \
\end{pmatrix}
\end{equation}

We can then evolve the initial state with the symplectic transformation $S=\text{e}^{\Omega H_C \sqrt{dt} }$: 

\begin{equation}
\begin{aligned}
\text{e}^{\Omega H_C \sqrt{dt}} \left( \mathbf{\sigma}_S\oplus\mathbf{\sigma}_B \right) \text{e}^{H_C \Omega^T \sqrt{dt}} 
= \mathbf{\sigma}_S\oplus\mathbf{\sigma}_B + \mathbf{\sigma}_{SB}\sqrt{dt} +\\ 
(A\boldsymbol{\sigma}_S + \boldsymbol{\sigma}_S A^T + D) \oplus (A\boldsymbol{\sigma}_B + \boldsymbol{\sigma}_B A^T + D)^T dt + ... \\
\Rightarrow \dot{\boldsymbol{\sigma}} = \mathbf{\sigma}_{SB}\sqrt{dt} +\\ 
(A\boldsymbol{\sigma}_S + \boldsymbol{\sigma}_S A^T + D) \oplus (A\boldsymbol{\sigma}_B + \boldsymbol{\sigma}_B A^T + D)^T dt + ...
\end{aligned}
\label{expansion}
\end{equation}
To obtain the equation for $\mathbf{\sigma}_S$, we can take the upper left block of the above equation in the first order in $dt$. Note that $\mathbf{\sigma}_{SB}$ has zero diagonal blocks, so it falls out \footnote{The upper right off-diagonal block is $\boldsymbol{\Omega} C \boldsymbol{\sigma}_B + \boldsymbol{\sigma}_S C \boldsymbol{\Omega}^T$}. We can now go back to denoting the system covariance matrix by $\mathbf{\sigma}$.

We assume that the initial state of the bath is the vacuum, so \( \boldsymbol{\sigma}_B = \mathbf{I} \). 
The steady state is obtained by setting \( \dot{\boldsymbol{\sigma}} = 0 \) in Eq.~\eqref{uncond}, but this unconditional evolution leads to a squeezing limit of 3 dB \cite{Collett1984, coherentfeedback} and cannot explain the higher entanglement observed in experiments.

This formulation is equivalent to the Langevin equation approach, as one can derive the equation of motion for the system operators in terms of $A, \ C$ and $\hat{H}_S$ \cite{serafinibook}: 

\begin{equation}
    \dot{\hat{r}}_k = i[\hat{H}_S, \hat{r}_k] + (A\tilde{r})_k + (\Omega C \tilde{r}_B)_k, 
\end{equation}
where $\tilde{r}$ $(\tilde{r}_B)$ denotes the vector of system (bath) operators.

\subsection{Conditional Dynamics and Continuous Monitoring}

A continuous measurement means that the measurement process is not instant and extracting information takes a non-zero amount of time \cite{Jacobs_2014}. It follows that as the measurement duration tends to zero, the extracted information must also go to zero. Thus, the state will be conditioned on the result of the measurement in the environment. Following the description of \cite{serafinibook}, the conditioning will result in the Gaussian channel on the first and second moments: 

\begin{equation}
\begin{aligned}
    \boldsymbol{\sigma} \rightarrow \boldsymbol{\sigma} - (\boldsymbol{\Omega} C \boldsymbol{\sigma}_B - \boldsymbol{\sigma} C \boldsymbol{\Omega}) (\boldsymbol{\sigma}_B + \boldsymbol{\sigma}_m)^{-1} (\cdots)^T\\
    \bar{r} \rightarrow \bar{r} + (\boldsymbol{\Omega} C \boldsymbol{\sigma}_B - \boldsymbol{\sigma} C \boldsymbol{\Omega}) (\boldsymbol{\sigma}_B + \boldsymbol{\sigma}_m)^{-1}. 
\end{aligned}
\label{monitormapo}
\end{equation}

The matrix $\boldsymbol{\Omega} C \boldsymbol{\sigma}_B - \boldsymbol{\sigma} C \boldsymbol{\Omega}$ is the inverse block diagonal matrix $\mathbf{\sigma}_{SB}$ in Eq. \ref{expansion}. It describes the built-up correlations during the system-environment interaction \cite{serafinibook}. Putting this into Eq. \ref{expansion}, we arrive at the \textit{Riccati equation}, which governs the time evolution of the covariance matrix under continuous measurement \cite{serafinibook, Genoni_2016, Generaldynemeasurements, Kansanen2025}:
\begin{equation}
\begin{aligned}
\dot{\boldsymbol{\sigma}} =\; & A \boldsymbol{\sigma} + \boldsymbol{\sigma} A^T + D \\
& - (\boldsymbol{\Omega} C \boldsymbol{\sigma}_B - \boldsymbol{\sigma} C \boldsymbol{\Omega}) (\boldsymbol{\sigma}_B + \boldsymbol{\sigma}_m)^{-1} (\cdots)^T, 
\end{aligned}
\label{riccati}
\end{equation}
where \( \boldsymbol{\sigma}_m \) characterizes the measurement that may contain noise. In steady state (\( \dot{\boldsymbol{\sigma}} = 0 \)), we obtain the \textit{algebraic Riccati equation}. For noisy heterodyne measurements, $\boldsymbol{\sigma}_m$ is given in Eq. \ref{effhet}. We can recover the unconditional dynamics by letting $\boldsymbol{\sigma}_m = x\mathbf{I}$ and taking the limit $x\rightarrow\infty$.

\subsection{Noise and Gaussian Channels}



Any Gaussian operation can be described by a completely positive (CP) Gaussian map \( \Phi \) with real matrices \( X \), \( Y \):
\begin{align}
\bar{r} &\rightarrow X\bar{r}, \\
\boldsymbol{\sigma} &\rightarrow X \boldsymbol{\sigma} X^T + Y,
\end{align}
subject to \( Y + i\boldsymbol{\Omega} \ge iX \boldsymbol{\Omega} X^T \). The \textit{dual map} \( \Phi^* \) introduces a transformation according to:
\begin{align}
X^* &= X^{-1}, \\
Y^* &= X^{-1} Y (X^{-1})^T,
\end{align}
which leads to an effective measurement matrix:
\begin{equation}
\boldsymbol{\sigma}_m \rightarrow X^* \boldsymbol{\sigma}_m X^{*T} + Y^*.
\label{noisemodel}
\end{equation}

To describe losses before measurement, we adopt the attenuator channel. It is described by the action of mixing the output modes with a thermal state with the covariance matrix $\bar{n}\mathbf{I} = (2N+1)\mathbf{I}$ on a beam-splitter \cite{gaussianquantumchannels, Generaldynemeasurements, serafinibook}: 

\begin{equation}
X = \cos\theta\, \mathbf{I}, \quad Y = \bar{n}\sin^2\theta\, \mathbf{I}, \quad 0 \le \theta \le \frac{\pi}{2},  \bar{n}\geq 1.
\end{equation}

The factor is the quadrature variance of the Gibbs state, which is a thermal state with $\langle \hat{c} \rangle = 0$ and $\langle \hat{c}^\dagger \hat{c} \rangle = N$ for each mode. For our definition of quadrature operators, we have $\langle \hat{x}^2 \rangle = \frac{1}{2}\langle \hat{c}^\dagger\hat{c} + \hat{c}\hat{c}^\dagger \rangle = \frac{1}{2}(2N+1) = \langle \hat{p}^2 \rangle$. Our covariance matrix is normalized to 1, so $\boldsymbol{\sigma}_{xx} = 2\langle \hat{x}^2 \rangle = (2N+1)$.



A quantum limited attenuator has $\bar{n}=1$. The effective heterodyne measurement becomes:
\begin{equation}
\boldsymbol{\sigma}_m = \mathbf{I} \left( \cos^{-2} \theta + \bar{n}\tan^2 \theta \right).
\label{effhet}
\end{equation}

The angle $\theta$ can be understood either as the efficiency of heterodyne measurements or as the losses that happen before the measurement. This model does not specify if they are due to internal dissipation or related to the measurement itself. The parameter $\bar{n}$ denotes the amount of thermal noise in the system. In the limit $\theta \rightarrow 0$, we recover the ideal heterodyne measurement matrix $\sigma_m = \mathbf{I}$. 

\subsection{Spontaneous parametric down-conversion}

Spontaneous parametric down-conversion in a 3-wave mixing regime is a process in which a photon $\gamma_p$ with energy $\Omega_p$ interacts with a nonlinear medium, causing the photon to split into two photons $\gamma_j$, $\gamma_k$, or signal and idler photons, with lower energies $\Omega_j$ and $\Omega_k$ while conserving energy with $\Omega_p=\Omega_j+\Omega_k$ and momentum: $k_p/2 - k_j = k_i - k_p/2$ \cite{PetrovninSPDC}. In each down-conversion process, the photons are created symmetrically around half of the pump frequency $\frac{\Omega_p}{2}$ \cite{PetrovninSPDC}. The Hamiltonian of the system describing the squeezing in a symmetric configuration of multiple pumps therefore consists of 2-mode squeezing terms: 

\begin{equation}
    \hat{S}_{jk,p} = i(\alpha_{p}\hat{a}_{j}^{\dagger}\hat{a}_{k}^{\dagger}-\alpha_{p}^{\ast}\hat{a}_{j}\hat{a}_{k}),
\end{equation}
where $\alpha_p = |\alpha_p| \text{e}^{i\phi_p}$ is the parametric coupling rate with phase $\phi_p$ associated with the amplitude $|\alpha_p|$ of a semiclassical pump operator that couples nonlinearly with the operators $\hat{a}_{j}^{\dagger}$ ($\hat{a}_{j}$)  and $\hat{a}_{k}^{\dagger}$ ($\hat{a}_{k}$), the creation and annihilation operators of the signal and idler modes. The index $p$ refers to the pump and $j,k$ to the two modes that are entangled by the pump tone. 

The full Hamiltonian of the system can be expressed compactly as an $\mathcal{H}$-graph state \cite{graphresource, graphtransrule, alocco2025programmablemicrowaveclusterstates}: 

\begin{equation}
    \hat{H}_S = \sum_{p} \sum_{\{jk\}_p} \hat{S}_{jk,p} = \frac{i}{2}\sum_{j, k} (\mathbf{A}_{jk} \hat{a}^\dagger_j \hat{a}^\dagger_k - \mathbf{A}^\ast_{jk} \hat{a}_j \hat{a}_k),
    \label{H-graph}
\end{equation}
where the first sum is computed on all pumps $p$ while the second sum counts on the mode pairs $j$, $k$ that are squeezed by the pump $p$ and $\mathbf{A}$ is the graph adjacency matrix, see Appendix \ref{adjacency matrix}. The graph is a visual representation of the interaction Hamiltonian and each edge symbolizes a squeezing channel inducing correlations.

In such a system, where each parametric pump creates mode pairs symmetrically around half the pump frequency, one can get an infinite number of squeezed pairs of modes. The achievable maximum number of modes can be estimated by $f_p/2\Delta_f$ where $\Delta_f$ is the finite bandwidth of a single mode. 

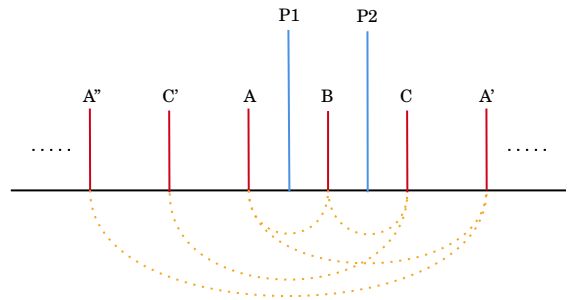
\begin{figure}[H]
\centering
\tikzset{every picture/.style={line width=0.75pt}} 

\begin{tikzpicture}[x=0.75pt,y=0.75pt,yscale=-1,xscale=1]

\draw    (190.13,190.32) -- (470.13,190.32) ;
\draw [color={rgb, 255:red, 74; green, 144; blue, 226 }  ,draw opacity=1 ][line width=0.75]    (330.13,109.32) -- (330.63,190.82) ;
\draw [color={rgb, 255:red, 208; green, 2; blue, 27 }  ,draw opacity=1 ]   (310,149) -- (310.13,190.25) ;
\draw [color={rgb, 255:red, 208; green, 2; blue, 27 }  ,draw opacity=1 ]   (350.13,150.25) -- (350.13,190.25) ;
\draw [color={rgb, 255:red, 245; green, 166; blue, 35 }  ,draw opacity=1 ] [dash pattern={on 0.84pt off 2.51pt}]  (310.13,190.25) .. controls (310,219.33) and (350.28,219.55) .. (350.13,190.25) ;
\draw [color={rgb, 255:red, 74; green, 144; blue, 226 }  ,draw opacity=1 ]   (370,110) -- (370.13,191.25) ;
\draw [color={rgb, 255:red, 245; green, 166; blue, 35 }  ,draw opacity=1 ] [dash pattern={on 0.84pt off 2.51pt}]  (350.13,190.25) .. controls (351,219.73) and (390,219.73) .. (390.13,191.25) ;
\draw [color={rgb, 255:red, 245; green, 166; blue, 35 }  ,draw opacity=1 ] [dash pattern={on 0.84pt off 2.51pt}]  (310.13,190.25) .. controls (310.28,239.55) and (430.15,239.43) .. (430,189.73) ;
\draw [color={rgb, 255:red, 208; green, 2; blue, 27 }  ,draw opacity=1 ]   (390,150) -- (390.13,191.25) ;
\draw [color={rgb, 255:red, 208; green, 2; blue, 27 }  ,draw opacity=1 ]   (430,149) -- (430.13,190.25) ;
\draw [color={rgb, 255:red, 208; green, 2; blue, 27 }  ,draw opacity=1 ]   (270,150) -- (270.13,191.25) ;
\draw [color={rgb, 255:red, 208; green, 2; blue, 27 }  ,draw opacity=1 ]   (230,149) -- (230.13,190.25) ;
\draw [color={rgb, 255:red, 245; green, 166; blue, 35 }  ,draw opacity=1 ] [dash pattern={on 0.84pt off 2.51pt}]  (270.27,191.77) .. controls (270,249.47) and (390,250.47) .. (390.13,191.25) ;
\draw [color={rgb, 255:red, 245; green, 166; blue, 35 }  ,draw opacity=1 ] [dash pattern={on 0.84pt off 2.51pt}]  (230.13,190.25) .. controls (230,261.47) and (430,261.47) .. (430.13,190.25) ;
\draw  [dash pattern={on 0.84pt off 2.51pt}]  (220,169.87) -- (200,169.87) ;
\draw  [dash pattern={on 0.84pt off 2.51pt}]  (460,169.87) -- (440,169.87) ;

\draw (323.85,96.77) node [anchor=north west][inner sep=0.75pt]  [font=\scriptsize] [align=left] {P1};
\draw (306.08,137.7) node [anchor=north west][inner sep=0.75pt]  [font=\scriptsize] [align=left] {A};
\draw (345.08,137.7) node [anchor=north west][inner sep=0.75pt]  [font=\scriptsize] [align=left] {B};
\draw (363.08,97.7) node [anchor=north west][inner sep=0.75pt]  [font=\scriptsize] [align=left] {P2};
\draw (385.08,137.7) node [anchor=north west][inner sep=0.75pt]  [font=\scriptsize] [align=left] {C};
\draw (425.08,137.7) node [anchor=north west][inner sep=0.75pt]  [font=\scriptsize] [align=left] {A'};
\draw (265.08,137.7) node [anchor=north west][inner sep=0.75pt]  [font=\scriptsize] [align=left] {C'};
\draw (225.08,137.7) node [anchor=north west][inner sep=0.75pt]  [font=\scriptsize] [align=left] {A''};
\end{tikzpicture}
\caption{Illustration of the mode-coupling sequence beyond the two-mode squeezed modes in the case of two parametric pumps. The dotted lines represent two-mode squeezing. We call these the first-order correlations between modes. However, because A and B are squeezed by P1, and B and C are squeezed by P2, there will be a beam-splitter correlation between A and C. Since these correlations grow proportionally to the product of two pump amplitudes, they are called second-order correlations between modes. These correlations are contained within the model and can be seen by expanding the unitary evolution operator generated by the Hamiltonian. In this figure, only first-order correlations are shown.}
\label{Fig:infinite idler sequence}
\end{figure}

The possible large number of modes is due to the fact that the frequency modes in the squeezed pairs are not unique: Any pair of frequencies that satisfy $\Omega_p=\Omega_j+\Omega_k$ will be squeezed. The result is that when you are entangling a pair of modes (A, B) to a TMS state with one pump and you add a second pump that squeezes another pair of modes (B, C), it will also squeeze mode A with a new mode A' that is equally apart in frequency but on the other side of that pump. Similarly, the first pump entangles modes C and A' symmetrically to additional modes C' and A'', respectively. The second pump will then entangle these modes to even more additional modes, and this mirroring sequence will continue until the device bandwidth is exceeded. The sequence described is illustrated in Fig. \ref{Fig:infinite idler sequence}. 

An important feature that we can infer from this mode-coupling sequence is that \textit{the placement of the second pump determines the mode structure of the symmetric system}; had we not added the second pump, we would only have had an infinite number of two-mode squeezed pairs and no coupling between different pairs. The second pump squeezes modes from different pairs with the same separation, resulting in a more involved structure. If we keep adding pumps with the same separation, we are only creating new couplings between the modes in this subset of frequencies. Thus, symmetric pumping produces a closed frequency subset, because no other frequencies satisfy energy conservation. 

\subsection{Squeezed mode pairs}
We place the modes on the real axis at positions \( x = \pm1, \pm3, \dots, 2n \pm 1 \), and the pump half-frequencies at \( x = 0, \pm2, \pm4, \dots, \pm2n \). A pump at position \( P \) squeezes all mode pairs symmetrically displaced from it:
\begin{equation}
    x = P \pm (2n+1), \quad n = 0, 1, 2, \dots
\end{equation}
These define the set of two-mode-squeezed (TMS) pairs (see Fig.~\ref{Fig:30 mode sym}).

To determine which modes are relevant for our calculations (i.e., those having first- or second-order correlations with modes $-1$ or $1$), we work up to 15 pumps, the furthest at \( P = \pm14 \). Solving
\begin{equation}
    P \pm (2n+1) = \pm1
\end{equation}
gives \( n = 7 \), corresponding to the pair \( (29, -1) \) and its mirror. Including both positive and negative sides, we count \(-29, -27, ..., 27, 29\), leading to a total of \textbf{30 directly correlated modes}.

To include beam-splitter (BS)–type second-order correlations, we find modes connected via an intermediate mode to $-1$ or $1$. Solving for TMS pairs that include $\pm29$ yields
\begin{equation}
    P \pm (2n+1) = \pm29,
\end{equation}
which gives \( n = 21 \), corresponding to the farthest relevant modes at \( \pm(2 \cdot 21 + 1) = \pm43 \). This adds 14 more modes (7 on each side), for a total system size of \textbf{44 modes}.

These formulas can describe any SPDC-based system: the expression for TMS pairs holds universally, while the specific bounds on \( P \) and \( x \) reflect our experimental constraints that limit the number of pumps to 15 and a truncated mode set.


\section{Graph theoretical considerations} \label{GraphTheory}

In this section, we detail the involved mode structures occurring in the devices exposed to multiple pumps and present a graph-theoretical tool for simplifying the problem of exactly solving the dynamics of a multi-mode Gaussian state.

\subsection{Symmetric pumping}
\begin{figure}[h]
\tikzset{every picture/.style={line width=0.5pt}} 

\begin{tikzpicture}[x=0.7pt,y=0.75pt,yscale=-1,xscale=1]
\draw    (161,110) -- (501,110) ;
\draw [color={rgb, 255:red, 208; green, 2; blue, 27 }  ,draw opacity=1 ][line width=0.75]    (330.13,27.32) -- (330.63,109.25) ;
\draw [draw opacity=1 ][line width=0.75][dash pattern={on 0.84pt off 2.51pt}]   (320,69) -- (320.13,109.25) ;
\draw [draw opacity=1 ][line width=0.75][dash pattern={on 0.84pt off 2.51pt}]   (340.13,69) -- (340.13,109.25) ;
\draw [draw opacity=1 ][line width=0.75][dash pattern={on 0.84pt off 2.51pt}]   (301,69) -- (301.13,109.25) ;
\draw [draw opacity=1 ][line width=0.75][dash pattern={on 0.84pt off 2.51pt}]   (360,69) -- (360.13,109.25) ;
\draw [draw opacity=1 ][line width=0.75]    (350.13,27.32) -- (350.63,108.82) ;
\draw [draw opacity=1 ][line width=0.75]    (310.13,27.32) -- (310.63,108.82) ;
\draw [draw opacity=1 ][line width=0.75][dash pattern={on 0.84pt off 2.51pt}]  (411,69) -- (411.13,109.25) ;
\draw [draw opacity=1 ][line width=0.75]    (420.13,27.32) -- (420.63,108.82) ; 
\draw [draw opacity=1 ][line width=0.75][dash pattern={on 0.84pt off 2.51pt}]   (431,69) -- (431.13,109.25) ;
\draw [draw opacity=1 ][line width=0.75][dash pattern={on 0.84pt off 2.51pt}]   (250,69) -- (250.13,109.25) ;
\draw [draw opacity=1 ][line width=0.75]    (239.13,27.32) -- (239.63,108.82) ; 
\draw [draw opacity=1 ][line width=0.75][dash pattern={on 0.84pt off 2.51pt}]   (230,69) -- (230.13,109.25) ;
\draw [draw opacity=1 ][line width=0.75]   (481,69) -- (481.13,109.25) ;
\draw [draw opacity=1 ][line width=0.75]   (180.72,69) -- (180.85,109.25) ;
\draw  [color={rgb, 255:red, 189; green, 16; blue, 224 }  ,draw opacity=1 ][dash pattern={on 0.84pt off 2.51pt}] (313.67,53) -- (346.67,53) -- (346.67,113) -- (313.67,113) -- cycle ;
\draw  [dash pattern={on 0.84pt off 2.51pt}]  (290,90) -- (262,90) ;
\draw  [dash pattern={on 0.84pt off 2.51pt}]  (398,90) -- (370,90) ;
\draw  [dash pattern={on 0.84pt off 2.51pt}]  (468,90) -- (440,90) ;
\draw  [dash pattern={on 0.84pt off 2.51pt}]  (220,90) -- (192,90) ;

\draw (323.85,12.77) node [anchor=north west][inner sep=0.75pt]  [font=\scriptsize] [align=left] {P0};
\draw (292.08,54.7) node [anchor=north west][inner sep=0.75pt]  [font=\scriptsize] [align=left] {\mbox{-}3};
\draw (314.08,54.7) node [anchor=north west][inner sep=0.75pt]  [font=\scriptsize] [align=left] {\mbox{-}1};
\draw (334.08,54.7) node [anchor=north west][inner sep=0.75pt]  [font=\scriptsize] [align=left] {1};
\draw (354.08,54.7) node [anchor=north west][inner sep=0.75pt]  [font=\scriptsize] [align=left] {3};
\draw (344.85,12.77) node [anchor=north west][inner sep=0.75pt]  [font=\scriptsize] [align=left] {P2};
\draw (301.85,12.77) node [anchor=north west][inner sep=0.75pt]  [font=\scriptsize] [align=left] {P-2};
\draw (405.08,54.7) node [anchor=north west][inner sep=0.75pt]  [font=\scriptsize] [align=left] {13};
\draw (411.85,12.77) node [anchor=north west][inner sep=0.75pt]  [font=\scriptsize] [align=left] {P14};
\draw (425.08,54.7) node [anchor=north west][inner sep=0.75pt]  [font=\scriptsize] [align=left] {15};
\draw (243.08,54.7) node [anchor=north west][inner sep=0.75pt]  [font=\scriptsize] [align=left] {\mbox{-}13};
\draw (230.85,12.77) node [anchor=north west][inner sep=0.75pt]  [font=\scriptsize] [align=left] {P-14};
\draw (222.08,54.7) node [anchor=north west][inner sep=0.75pt]  [font=\scriptsize] [align=left] {\mbox{-}15};
\draw (474.08,54.7) node [anchor=north west][inner sep=0.75pt]  [font=\scriptsize] [align=left] {43};
\draw (173.8,54.7) node [anchor=north west][inner sep=0.75pt]  [font=\scriptsize] [align=left] {\mbox{-}43};

\draw (315,120) node [anchor=north west][inner sep=0.75pt]  [font=\scriptsize] [align=left] {$\displaystyle \Delta$};
\draw (336,120) node [anchor=north west][inner sep=0.75pt]  [font=\scriptsize] [align=left] {$\displaystyle \Delta$};

\draw (313,116.23) -- (313,119.83);
\draw (313,119.83) -- (327,119.83);
\draw (327,119.83) -- (327,116.23);

\draw (334,116.23) -- (334,119.83);
\draw (334,119.83) -- (348,119.83);
\draw (348,119.83) -- (348,116.23);

\end{tikzpicture}

\begin{tikzpicture}[scale=2, every node/.style={circle, draw, inner sep=1pt, font=\scriptsize}]
\def\n{30}
\def\r{2}

\foreach \i/\lab in {
0/-15, 1/-14, 2/-13, 3/-12, 4/-11, 5/-10, 6/-9, 7/-8, 8/-7, 9/-6,
10/-5, 11/-4, 12/-3, 13/-2, 14/-1,
15/1, 16/2, 17/3, 18/4, 19/5, 20/6, 21/7, 22/8, 23/9, 24/10,
25/11, 26/12, 27/13, 28/14, 29/15
} {
    \pgfmathsetmacro{\angle}{90 - 360*\i/\n}
    
    \pgfmathtruncatemacro{\display}{%
        ifthenelse(\lab<0, 2*\lab+1, 2*\lab-1)
    }
    
    \node (v\i) at (\angle:\r) {\display};
}


\foreach \a/\b in {
-1/1, -3/3, -5/5, -7/7, -9/9, -11/11, -13/13, -15/15, -17/17, -19/19, -21/21, -23/23, -25/25, -27/27, -29/29
}{
    \pgfmathtruncatemacro{\ia}{\a<0 ? int(\a/2 + 15) : int((\a+1)/2 + 14)}
    \pgfmathtruncatemacro{\ib}{\b<0 ? int(\b/2 + 15) : int((\b+1)/2 + 14)}
    \draw[red, thick] (v\ia) -- (v\ib);
}

\foreach \a/\b in {
1/3, -1/5, -3/7, -5/9, -7/11, -9/13, -11/15, -13/17, -15/19, -17/21, -19/23, -21/25, -23/27, -25/29
}{
    \pgfmathtruncatemacro{\ia}{\a<0 ? int(\a/2 + 15) : int((\a+1)/2 + 14)}
    \pgfmathtruncatemacro{\ib}{\b<0 ? int(\b/2 + 15) : int((\b+1)/2 + 14)}
    \draw (v\ia) -- (v\ib);
}

\foreach \a/\b in {
-1/-3, 1/-5, 3/-7, 5/-9, 7/-11, 9/-13, 11/-15, 13/-17, 15/-19, 17/-21, 19/-23, 21/-25, 23/-27, 25/-29
}{
    \pgfmathtruncatemacro{\ia}{\a<0 ? int(\a/2 + 15) : int((\a+1)/2 + 14)}
    \pgfmathtruncatemacro{\ib}{\b<0 ? int(\b/2 + 15) : int((\b+1)/2 + 14)}
    \draw (v\ia) -- (v\ib);
}

\foreach \a/\b in {
3/5, 1/7, -1/9, -3/11, -5/13, -7/15, -9/17, -11/19, -13/21, -15/23, -17/25, -19/27, -21/29
}{
    \pgfmathtruncatemacro{\ia}{\a<0 ? int(\a/2 + 15) : int((\a+1)/2 + 14)}
    \pgfmathtruncatemacro{\ib}{\b<0 ? int(\b/2 + 15) : int((\b+1)/2 + 14)}
    \draw (v\ia) -- (v\ib);
}

\foreach \a/\b in {
-3/-5, -1/-7, 1/-9, 3/-11, 5/-13, 7/-15, 9/-17, 11/-19, 13/-21, 15/-23, 17/-25, 19/-27, 21/-29
}{
    \pgfmathtruncatemacro{\ia}{\a<0 ? int(\a/2 + 15) : int((\a+1)/2 + 14)}
    \pgfmathtruncatemacro{\ib}{\b<0 ? int(\b/2 + 15) : int((\b+1)/2 + 14)}
    \draw (v\ia) -- (v\ib);
}

\foreach \a/\b in {
5/7, 3/9, 1/11, -1/13, -3/15, -5/17, -7/19, -9/21, -11/23, -13/25, -15/27, -17/29
}{
    \pgfmathtruncatemacro{\ia}{\a<0 ? int(\a/2 + 15) : int((\a+1)/2 + 14)}
    \pgfmathtruncatemacro{\ib}{\b<0 ? int(\b/2 + 15) : int((\b+1)/2 + 14)}
    \draw (v\ia) -- (v\ib);
}

\foreach \a/\b in {
-5/-7, -3/-9, -1/-11, 1/-13, 3/-15, 5/-17, 7/-19, 9/-21, 11/-23, 13/-25, 15/-27, 17/-29
}{
    \pgfmathtruncatemacro{\ia}{\a<0 ? int(\a/2 + 15) : int((\a+1)/2 + 14)}
    \pgfmathtruncatemacro{\ib}{\b<0 ? int(\b/2 + 15) : int((\b+1)/2 + 14)}
    \draw (v\ia) -- (v\ib);
}

\foreach \a/\b in {
7/9, 5/11, 3/13, 1/15, -1/17, -3/19, -5/21, -7/23, -9/25, -11/27, -13/29
}{
    \pgfmathtruncatemacro{\ia}{\a<0 ? int(\a/2 + 15) : int((\a+1)/2 + 14)}
    \pgfmathtruncatemacro{\ib}{\b<0 ? int(\b/2 + 15) : int((\b+1)/2 + 14)}
    \draw (v\ia) -- (v\ib);
}

\foreach \a/\b in {
-7/-9, -5/-11, -3/-13, -1/-15, 1/-17, 3/-19, 5/-21, 7/-23, 9/-25, 11/-27, 13/-29
}{
    \pgfmathtruncatemacro{\ia}{\a<0 ? int(\a/2 + 15) : int((\a+1)/2 + 14)}
    \pgfmathtruncatemacro{\ib}{\b<0 ? int(\b/2 + 15) : int((\b+1)/2 + 14)}
    \draw (v\ia) -- (v\ib);
}

\foreach \a/\b in {
9/11, 7/13, 5/15, 3/17, 1/19, -1/21, -3/23, -5/25, -7/27, -9/29
}{
    \pgfmathtruncatemacro{\ia}{\a<0 ? int(\a/2 + 15) : int((\a+1)/2 + 14)}
    \pgfmathtruncatemacro{\ib}{\b<0 ? int(\b/2 + 15) : int((\b+1)/2 + 14)}
    \draw (v\ia) -- (v\ib);
}

\foreach \a/\b in {
-9/-11, -7/-13, -5/-15, -3/-17, -1/-19, 1/-21, 3/-23, 5/-25, 7/-27, 9/-29
}{
    \pgfmathtruncatemacro{\ia}{\a<0 ? int(\a/2 + 15) : int((\a+1)/2 + 14)}
    \pgfmathtruncatemacro{\ib}{\b<0 ? int(\b/2 + 15) : int((\b+1)/2 + 14)}
    \draw (v\ia) -- (v\ib);
}

\foreach \a/\b in {
11/13, 9/15, 7/17, 5/19, 3/21, 1/23, -1/25, -3/27, -5/29
}{
    \pgfmathtruncatemacro{\ia}{\a<0 ? int(\a/2 + 15) : int((\a+1)/2 + 14)}
    \pgfmathtruncatemacro{\ib}{\b<0 ? int(\b/2 + 15) : int((\b+1)/2 + 14)}
    \draw (v\ia) -- (v\ib);
}

\foreach \a/\b in {
-11/-13, -9/-15, -7/-17, -5/-19, -3/-21, -1/-23, 1/-25, 3/-27, 5/-29
}{
    \pgfmathtruncatemacro{\ia}{\a<0 ? int(\a/2 + 15) : int((\a+1)/2 + 14)}
    \pgfmathtruncatemacro{\ib}{\b<0 ? int(\b/2 + 15) : int((\b+1)/2 + 14)}
    \draw (v\ia) -- (v\ib);
}

\foreach \a/\b in {
13/15, 11/17, 9/19, 7/21, 5/23, 3/25, 1/27, -1/29
}{
    \pgfmathtruncatemacro{\ia}{\a<0 ? int(\a/2 + 15) : int((\a+1)/2 + 14)}
    \pgfmathtruncatemacro{\ib}{\b<0 ? int(\b/2 + 15) : int((\b+1)/2 + 14)}
    \draw (v\ia) -- (v\ib);
}

\foreach \a/\b in {
-13/-15, -11/-17, -9/-19, -7/-21, -5/-23, -3/-25, -1/-27, 1/-29
}{
    \pgfmathtruncatemacro{\ia}{\a<0 ? int(\a/2 + 15) : int((\a+1)/2 + 14)}
    \pgfmathtruncatemacro{\ib}{\b<0 ? int(\b/2 + 15) : int((\b+1)/2 + 14)}
    \draw (v\ia) -- (v\ib);
}

\end{tikzpicture}

\caption{Mode frequencies marked by $\pm(2m+1)$, for $m=0, ..., 21$ and a 30-mode subgraph (contains modes denoted by dashed lines) constructed for a system with 15 symmetrically placed pumps. We are investigating the bipartite entanglement of the pair (-1, 1), inside the dashed square. Each edge represents a TMS term in the Hamiltonian. The weight for each edge is specified by the corresponding classical pump amplitude. Edges generated by P0 are indicated by red. We can see that the center pair (-1, 1) has 15 neighbors while the furthest pairs $\pm 27, \pm29$ have only 8 neighbors. This shows that as we traverse the graph further away from the center pair (-1, 1), each new added mode will not entangle as much to the system as the modes closer to the center.}

\label{Fig:30 mode sym}
\end{figure}
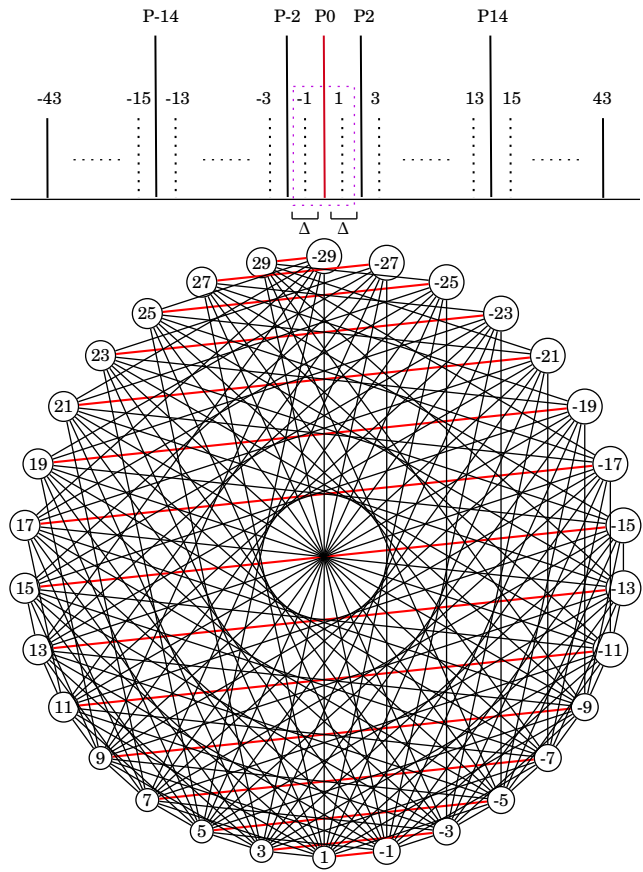

\noindent Consider a system of 44 equally spaced modes separated by $\Delta$, coupled together by 15 pumps with equal spacing of $2\Delta$ (half-pump frequencies spaced by $\Delta$) as illustrated in Fig. \ref{Fig:30 mode sym}. Let us refer to the modes of the symmetric system as the \textit{system modes}. 

We can then draw a graph where each node represents a mode, each weight represents a squeezing amplitude (proportional to the pump amplitude connecting the modes), and each edge represents a squeezing term in the Hamiltonian. Similarly, we can computer-generate the adjacency matrix given a graph (see Appendix \ref{adjacency matrix}) \cite{handbook}. The adjacency matrix (denoted "$\mathbf{A}$") is important because it turns out that the Hamiltonian matrix has the explicit form, as can be seen from Eq. \ref{H-graph}: 

\begin{equation}
     H_S=i\left(\begin{matrix}
\mathbf{A}&0\\
0&-\mathbf{A}^\ast
\end{matrix}\right)
\end{equation}
where $()^\ast$ represents element-wise complex conjugation. We can then use this Hamiltonian matrix to solve the steady-state Eq. (\ref{riccati}). Note that this is in the $\{\hat{a}^\dagger_{-1}, \hat{a}^\dagger_{1}, \hat{a}^\dagger_{-3}, \hat{a}^\dagger_{3}, ..., \hat{a}_{-1}, \hat{a}_{1}, \hat{a}_{-3}, \hat{a}_{3}...\}$ mode basis, so we need to transform to the quadrature basis $\{\hat{x}_{-1}, \hat{p}_{-1}, \hat{x}_{1}, \hat{p}_{1}...\}$ when we want to evaluate the correlations between quadratures of different modes. 

\subsection{Asymmetric pumping}

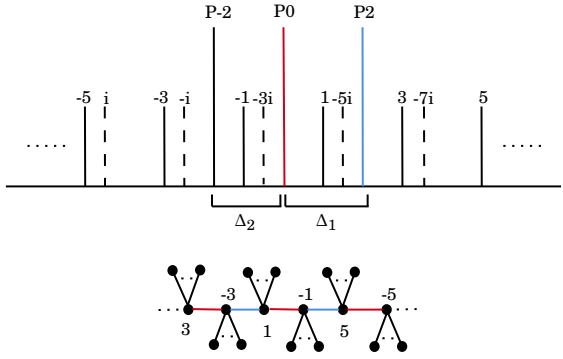
\begin{figure}[tb]
\tikzset{every picture/.style={line width=0.75pt}} 

\begin{tikzpicture}[x=0.75pt,y=0.75pt,yscale=-1,xscale=1]
\draw    (210.13,122.32) -- (490.13,122.32) ;
\draw [color={rgb, 255:red, 208; green, 2; blue, 27 }  ,draw opacity=1 ][line width=0.75]    (350.13,42) -- (350.63,122.25) ;
\draw [draw opacity=1 ]   (330,82) -- (330.13,122.25) ;
\draw [draw opacity=1 ]   (370.13,82) -- (370.13,122.25) ;
\draw [color={rgb, 255:red, 74; green, 144; blue, 226 }  ,draw opacity=1 ]   (390,42) -- (390.13,122.25) ;
\draw [draw opacity=1 ]   (410,82) -- (410.13,122.25) ;
\draw [draw opacity=1 ]   (450,82) -- (450.13,122.25) ;
\draw [draw opacity=1 ]   (290,82) -- (290.13,122.25) ;
\draw [draw opacity=1 ]   (250,82) -- (250.13,122.25) ;
\draw  [dash pattern={on 0.84pt off 2.51pt}]  (240,101.87) -- (220,101.87) ;
\draw  [dash pattern={on 0.84pt off 2.51pt}]  (480,101.87) -- (460,101.87) ;
\draw [draw opacity=1 ]   (315,42) -- (315.13,122.25) ;
\draw    (314.08,132.4) -- (349.08,132.4) ;
\draw    (314.08,132.4) -- (314.08,126.4) ;
\draw    (348.58,132.4) -- (348.58,126.4) ;
\draw    (351.08,132.4) -- (393.08,132.4) ;
\draw    (351.08,132.4) -- (351.08,126.4) ;
\draw    (392.58,132.4) -- (392.58,126.4) ;
\draw [draw opacity=1 ] [dash pattern={on 4.5pt off 4.5pt}]  (300,82) -- (300.13,122.25) ;
\draw [draw opacity=1 ] [dash pattern={on 4.5pt off 4.5pt}]  (260,82) -- (260.13,122.25) ;
\draw [draw opacity=1 ] [dash pattern={on 4.5pt off 4.5pt}]  (340,82) -- (340.13,121.25) ;
\draw [draw opacity=1 ] [dash pattern={on 4.5pt off 4.5pt}]  (380,82) -- (380.13,122.25) ;
\draw [draw opacity=1 ] [dash pattern={on 4.5pt off 4.5pt}]  (421,82) -- (421.13,122.25) ;
\draw  [fill={rgb, 255:red, 0; green, 0; blue, 0 }  ,fill opacity=1 ] (319.08,184.5) .. controls (319.08,183.26) and (320.02,182.25) .. (321.17,182.25) .. controls (322.32,182.25) and (323.25,183.26) .. (323.25,184.5) .. controls (323.25,185.74) and (322.32,186.75) .. (321.17,186.75) .. controls (320.02,186.75) and (319.08,185.74) .. (319.08,184.5) -- cycle ;
\draw  [fill={rgb, 255:red, 0; green, 0; blue, 0 }  ,fill opacity=1 ] (338.25,184.5) .. controls (338.25,183.26) and (339.18,182.25) .. (340.33,182.25) .. controls (341.48,182.25) and (342.42,183.26) .. (342.42,184.5) .. controls (342.42,185.74) and (341.48,186.75) .. (340.33,186.75) .. controls (339.18,186.75) and (338.25,185.74) .. (338.25,184.5) -- cycle ;
\draw  [fill={rgb, 255:red, 0; green, 0; blue, 0 }  ,fill opacity=1 ] (400.08,184.5) .. controls (400.08,183.26) and (401.02,182.25) .. (402.17,182.25) .. controls (403.32,182.25) and (404.25,183.26) .. (404.25,184.5) .. controls (404.25,185.74) and (403.32,186.75) .. (402.17,186.75) .. controls (401.02,186.75) and (400.08,185.74) .. (400.08,184.5) -- cycle ;
\draw  [fill={rgb, 255:red, 0; green, 0; blue, 0 }  ,fill opacity=1 ] (300.08,184.5) .. controls (300.08,183.26) and (301.02,182.25) .. (302.17,182.25) .. controls (303.32,182.25) and (304.25,183.26) .. (304.25,184.5) .. controls (304.25,185.74) and (303.32,186.75) .. (302.17,186.75) .. controls (301.02,186.75) and (300.08,185.74) .. (300.08,184.5) -- cycle ;
\draw  [fill={rgb, 255:red, 0; green, 0; blue, 0 }  ,fill opacity=1 ] (358.08,184.5) .. controls (358.08,183.26) and (359.02,182.25) .. (360.17,182.25) .. controls (361.32,182.25) and (362.25,183.26) .. (362.25,184.5) .. controls (362.25,185.74) and (361.32,186.75) .. (360.17,186.75) .. controls (359.02,186.75) and (358.08,185.74) .. (358.08,184.5) -- cycle ;
\draw  [fill={rgb, 255:red, 0; green, 0; blue, 0 }  ,fill opacity=1 ] (378.08,184.5) .. controls (378.08,183.26) and (379.02,182.25) .. (380.17,182.25) .. controls (381.32,182.25) and (382.25,183.26) .. (382.25,184.5) .. controls (382.25,185.74) and (381.32,186.75) .. (380.17,186.75) .. controls (379.02,186.75) and (378.08,185.74) .. (378.08,184.5) -- cycle ;
\draw [color={rgb, 255:red, 208; green, 2; blue, 27 }  ,draw opacity=1 ]   (319.08,184.5) -- (311.05,184.36) -- (304.15,184.23) ;
\draw [color={rgb, 255:red, 208; green, 2; blue, 27 }  ,draw opacity=1 ]   (400.08,184.5) -- (382.17,184.5) ;
\draw [color={rgb, 255:red, 74; green, 144; blue, 226 }  ,draw opacity=1 ]   (323.25,184.5) -- (338.25,184.5) ;
\draw [color={rgb, 255:red, 74; green, 144; blue, 226 }  ,draw opacity=1 ]   (362.17,184.5) -- (378.17,184.5) ;
\draw [color={rgb, 255:red, 208; green, 2; blue, 27 }  ,draw opacity=1 ]   (358.08,184.5) -- (350.05,184.36) -- (343.15,184.23) ;
\draw  [dash pattern={on 0.84pt off 2.51pt}]  (297.17,184.5) -- (284.1,184.77) ;
\draw  [dash pattern={on 0.84pt off 2.51pt}]  (417.32,184.23) -- (404.25,184.5) ;
\draw    (294.1,164.77) -- (302.1,182.77) ;
\draw  [dash pattern={on 0.84pt off 2.51pt}]  (298.6,168.77) -- (305.6,168.77) ;
\draw    (301.08,184.5) -- (308.1,164.77) ;
\draw  [fill={rgb, 255:red, 0; green, 0; blue, 0 }  ,fill opacity=1 ] (306.02,164.52) .. controls (306.02,163.27) and (306.95,162.27) .. (308.1,162.27) .. controls (309.25,162.27) and (310.18,163.27) .. (310.18,164.52) .. controls (310.18,165.76) and (309.25,166.77) .. (308.1,166.77) .. controls (306.95,166.77) and (306.02,165.76) .. (306.02,164.52) -- cycle ;
\draw  [fill={rgb, 255:red, 0; green, 0; blue, 0 }  ,fill opacity=1 ] (292.1,164.77) .. controls (292.1,163.52) and (293.03,162.52) .. (294.18,162.52) .. controls (295.33,162.52) and (296.27,163.52) .. (296.27,164.77) .. controls (296.27,166.01) and (295.33,167.02) .. (294.18,167.02) .. controls (293.03,167.02) and (292.1,166.01) .. (292.1,164.77) -- cycle ;
\draw    (329.05,202.11) -- (321.33,183.99) ;
\draw  [dash pattern={on 0.84pt off 2.51pt}]  (324.61,198.04) -- (317.61,197.93) ;
\draw    (322.38,182.27) -- (315.05,201.89) ;
\draw  [fill={rgb, 255:red, 0; green, 0; blue, 0 }  ,fill opacity=1 ] (317.13,202.17) .. controls (317.11,203.41) and (316.16,204.41) .. (315.01,204.39) .. controls (313.86,204.37) and (312.94,203.35) .. (312.96,202.11) .. controls (312.98,200.86) and (313.93,199.87) .. (315.08,199.89) .. controls (316.23,199.91) and (317.15,200.93) .. (317.13,202.17) -- cycle ;
\draw  [fill={rgb, 255:red, 0; green, 0; blue, 0 }  ,fill opacity=1 ] (331.05,202.14) .. controls (331.03,203.38) and (330.08,204.38) .. (328.93,204.36) .. controls (327.78,204.34) and (326.86,203.32) .. (326.88,202.08) .. controls (326.9,200.83) and (327.85,199.84) .. (329,199.86) .. controls (330.15,199.88) and (331.07,200.9) .. (331.05,202.14) -- cycle ;
\draw    (332.1,165.77) -- (340.1,183.77) ;
\draw  [dash pattern={on 0.84pt off 2.51pt}]  (336.6,169.77) -- (343.6,169.77) ;
\draw    (339.08,185.5) -- (346.1,165.77) ;
\draw  [fill={rgb, 255:red, 0; green, 0; blue, 0 }  ,fill opacity=1 ] (344.02,165.52) .. controls (344.02,164.27) and (344.95,163.27) .. (346.1,163.27) .. controls (347.25,163.27) and (348.18,164.27) .. (348.18,165.52) .. controls (348.18,166.76) and (347.25,167.77) .. (346.1,167.77) .. controls (344.95,167.77) and (344.02,166.76) .. (344.02,165.52) -- cycle ;
\draw  [fill={rgb, 255:red, 0; green, 0; blue, 0 }  ,fill opacity=1 ] (330.1,165.77) .. controls (330.1,164.52) and (331.03,163.52) .. (332.18,163.52) .. controls (333.33,163.52) and (334.27,164.52) .. (334.27,165.77) .. controls (334.27,167.01) and (333.33,168.02) .. (332.18,168.02) .. controls (331.03,168.02) and (330.1,167.01) .. (330.1,165.77) -- cycle ;
\draw    (368.05,203.11) -- (360.33,184.99) ;
\draw  [dash pattern={on 0.84pt off 2.51pt}]  (363.61,199.04) -- (356.61,198.93) ;
\draw    (361.38,183.27) -- (354.05,202.89) ;
\draw  [fill={rgb, 255:red, 0; green, 0; blue, 0 }  ,fill opacity=1 ] (356.13,203.17) .. controls (356.11,204.41) and (355.16,205.41) .. (354.01,205.39) .. controls (352.86,205.37) and (351.94,204.35) .. (351.96,203.11) .. controls (351.98,201.86) and (352.93,200.87) .. (354.08,200.89) .. controls (355.23,200.91) and (356.15,201.93) .. (356.13,203.17) -- cycle ;
\draw  [fill={rgb, 255:red, 0; green, 0; blue, 0 }  ,fill opacity=1 ] (370.05,203.14) .. controls (370.03,204.38) and (369.08,205.38) .. (367.93,205.36) .. controls (366.78,205.34) and (365.86,204.32) .. (365.88,203.08) .. controls (365.9,201.83) and (366.85,200.84) .. (368,200.86) .. controls (369.15,200.88) and (370.07,201.9) .. (370.05,203.14) -- cycle ;
\draw    (372.1,165.77) -- (380.1,183.77) ;
\draw  [dash pattern={on 0.84pt off 2.51pt}]  (376.6,169.77) -- (383.6,169.77) ;
\draw    (379.08,185.5) -- (386.1,165.77) ;
\draw  [fill={rgb, 255:red, 0; green, 0; blue, 0 }  ,fill opacity=1 ] (384.02,165.52) .. controls (384.02,164.27) and (384.95,163.27) .. (386.1,163.27) .. controls (387.25,163.27) and (388.18,164.27) .. (388.18,165.52) .. controls (388.18,166.76) and (387.25,167.77) .. (386.1,167.77) .. controls (384.95,167.77) and (384.02,166.76) .. (384.02,165.52) -- cycle ;
\draw  [fill={rgb, 255:red, 0; green, 0; blue, 0 }  ,fill opacity=1 ] (370.1,165.77) .. controls (370.1,164.52) and (371.03,163.52) .. (372.18,163.52) .. controls (373.33,163.52) and (374.27,164.52) .. (374.27,165.77) .. controls (374.27,167.01) and (373.33,168.02) .. (372.18,168.02) .. controls (371.03,168.02) and (370.1,167.01) .. (370.1,165.77) -- cycle ;
\draw    (410.05,203.11) -- (402.33,184.99) ;
\draw  [dash pattern={on 0.84pt off 2.51pt}]  (405.61,199.04) -- (398.61,198.93) ;
\draw    (403.38,183.27) -- (396.05,202.89) ;
\draw  [fill={rgb, 255:red, 0; green, 0; blue, 0 }  ,fill opacity=1 ] (398.13,203.17) .. controls (398.11,204.41) and (397.16,205.41) .. (396.01,205.39) .. controls (394.86,205.37) and (393.94,204.35) .. (393.96,203.11) .. controls (393.98,201.86) and (394.93,200.87) .. (396.08,200.89) .. controls (397.23,200.91) and (398.15,201.93) .. (398.13,203.17) -- cycle ;
\draw  [fill={rgb, 255:red, 0; green, 0; blue, 0 }  ,fill opacity=1 ] (412.05,203.14) .. controls (412.03,204.38) and (411.08,205.38) .. (409.93,205.36) .. controls (408.78,205.34) and (407.86,204.32) .. (407.88,203.08) .. controls (407.9,201.83) and (408.85,200.84) .. (410,200.86) .. controls (411.15,200.88) and (412.07,201.9) .. (412.05,203.14) -- cycle ;

\draw (343.85,28.77) node [anchor=north west][inner sep=0.75pt]  [font=\scriptsize] [align=left] {P0};
\draw (324.08,72.7) node [anchor=north west][inner sep=0.75pt]  [font=\scriptsize] [align=left] {\mbox{-}1};
\draw (367.08,72.7) node [anchor=north west][inner sep=0.75pt]  [font=\scriptsize] [align=left] {1};
\draw (384.08,29.7) node [anchor=north west][inner sep=0.75pt]  [font=\scriptsize] [align=left] {P2};
\draw (406.08,72.7) node [anchor=north west][inner sep=0.75pt]  [font=\scriptsize] [align=left] {3};
\draw (447.08,72.7) node [anchor=north west][inner sep=0.75pt]  [font=\scriptsize] [align=left] {5};
\draw (283.08,72.7) node [anchor=north west][inner sep=0.75pt]  [font=\scriptsize] [align=left] {\mbox{-}3};
\draw (244.08,72.7) node [anchor=north west][inner sep=0.75pt]  [font=\scriptsize] [align=left] {\mbox{-}5};
\draw (309.08,29.7) node [anchor=north west][inner sep=0.75pt]  [font=\scriptsize] [align=left] {P-2};
\draw (324,135) node [anchor=north west][inner sep=0.75pt]  [font=\scriptsize] [align=left] {$\displaystyle \Delta _{2}$};
\draw (365,135) node [anchor=north west][inner sep=0.75pt]  [font=\scriptsize] [align=left] {$\displaystyle \Delta _{1}$};
\draw (296.08,72.7) node [anchor=north west][inner sep=0.75pt]  [font=\scriptsize] [align=left] {\mbox{-}i};
\draw (258.08,72.7) node [anchor=north west][inner sep=0.75pt]  [font=\scriptsize] [align=left] {i};
\draw (333.08,72.7) node [anchor=north west][inner sep=0.75pt]  [font=\scriptsize] [align=left] {\mbox{-}3i};
\draw (373.08,72.7) node [anchor=north west][inner sep=0.75pt]  [font=\scriptsize] [align=left] {\mbox{-}5i};
\draw (414.08,72.7) node [anchor=north west][inner sep=0.75pt]  [font=\scriptsize] [align=left] {\mbox{-}7i};
\draw (297,190) node [anchor=north west][inner sep=0.75pt]  [font=\scriptsize] [align=left] {3};
\draw (338,191) node [anchor=north west][inner sep=0.75pt]  [font=\scriptsize] [align=left] {1};
\draw (398,172) node [anchor=north west][inner sep=0.75pt]  [font=\scriptsize] [align=left] {\mbox{-}5};
\draw (317,172) node [anchor=north west][inner sep=0.75pt]  [font=\scriptsize] [align=left] {\mbox{-}3};
\draw (356,172) node [anchor=north west][inner sep=0.75pt]  [font=\scriptsize] [align=left] {\mbox{-}1};
\draw (377,191) node [anchor=north west][inner sep=0.75pt]  [font=\scriptsize] [align=left] {5};

\end{tikzpicture}

\caption{Building up the asymmetric pump configuration: The basic TMS entanglement sequence is produced by pumps P0 and P2, with their half frequencies denoted by red (P0) and blue (P2) lines. The third pump is placed so that the distance $\Delta_2$ to P0 is different from $\Delta_1$ between P0 and P2. The dashed lines illustrate the sequence of P-2 added frequencies "Mi" with M referring to the mode frequencies established by pumps P0 and P2: extra mode frequency Mi is entangled with mode M. Below is the graphical description of a system with asymmetric pumps. Red edges represent TMS by P0, blue edges represent TMS by pump P2 and black edges represent TMS by subsequent asymmetric pumps.}
\label{Fig:asymmetric pumps}
\end{figure}

When pumps are placed asymmetrically, each with a unique displacement relative to the first, the system quickly becomes complex. However, we simplify the setup by assuming that the first two pumps generate a fixed set of 44 modes as in Fig.~\ref{Fig:30 mode sym}, called the \textit{system modes}. Starting from the third pump, each additional pump generates a new system of 44 modes, which we refer to as \textit{extra modes}. These systems do not couple to each other as the pumps were displaced asymmetrically. To keep the problem tractable, we ignore additional second-order correlations that are generated by the pre-existing pumps. Thus, each pump beyond the second introduces 44 modes per pump to the calculation, illustrated in Fig. \ref{Fig:asymmetric pumps}.

For example, the third pump (P--2) generates 44 extra modes: the modes to the right of P0 couple to modes $-i, -3i, -5i, \dots$, and those to the left to $i, 3i, 5i, \dots$. Since these modes do not couple among themselves due to the asymmetric configuration, the adjacency matrix for a 3-pump system takes the form:
\begin{equation}
    \mathbf{A} =
    \begin{pmatrix}
        \text{M}_{2p} & \alpha_3 \mathbf{I} \\
        \alpha_3 \mathbf{I} & 0
    \end{pmatrix}
\end{equation}
where $\text{M}_{2p}$ is the $44 \times 44$ adjacency matrix for the symmetric two-pump system, $\mathbf{I}$ the identity matrix, and $\alpha_3$ the amplitude of the third pump.

Adding more asymmetric pumps follows the same pattern. Each new extra mode layer connects only to the system modes. The resulting block structure generalizes to:
\begin{equation}
    \mathbf{A} =
    \begin{pmatrix}
        \text{M}_{2p} & \alpha_3 \mathbf{I} & \alpha_4 \mathbf{I} & \dots & \alpha_n \mathbf{I} \\
        \alpha_3 \mathbf{I} & 0 & 0 & \dots & 0 \\
        \alpha_4 \mathbf{I} & 0 & 0 & \dots & 0 \\
        \vdots & \vdots & \vdots & \ddots & \vdots \\
        \alpha_n \mathbf{I} & 0 & 0 & \dots & 0
    \end{pmatrix}
    \label{asymgraph}
\end{equation}

Here, each new block row/column corresponds to 44 extra modes introduced by an additional asymmetric pump. The resulting graph is a central line of system modes (from pumps 1 and 2), each connected to $n{-}2$ additional extra modes.

In the symmetric configuration, the number of relevant frequencies to the calculation and experiment stays relatively low, and the correlation structure, represented visually by a graph, stays sufficiently simple. With this type of configuration, the effects of pump power and coupling topology are made more intuitive. Also, as correlations are only created inside a fixed mode set, we get much more second-order BS-type correlations. In that case, the relative phases that determine the signs of the squeezing interactions sometimes lead to destructive interference. In the asymmetric configuration, the correlation structure is much more complicated, as every pump adds many new idler modes, which do not couple to each other. Thus, the total number of modes that have at least a second-order correlation with the center-most pair is much larger than in the symmetric case.

The symmetric pump configuration has a mirror symmetry only if there are the same number of pumps placed in equal intervals on each side of the first pump. It is equivalent to exchanging all labels with a minus sign and a positive sign. The graph does not have translation symmetry, because the modes that are closer to the center have more edges as indicated by Fig. \ref{Fig:30 mode sym}. As the graph is finite, traversing the graph in any direction means moving away from or toward the center-most pair, and their distance from the center determines how significant they are for the bipartite entanglement reduction of the principal pair. In other words, every mode and its mirror counterpart are identical and a unique pair. As the number of pumps increases, every mode is brought closer to the center-most pair via edges in the graph. As noted before, we expect this to redistribute the logarithmic negativity to the network. On the other hand, the asymmetric pump configuration does not even have mirror symmetry because the pumps are placed on irregular intervals. These symmetries have no observable effect for our purposes. 

\section{Results and Discussion} \label{ResultsDisc}

\subsection{Theory}
\begin{figure}
    \centering
    \begin{tikzpicture}
        \node[anchor=north west] (img1) at (-7.5,0) {\includegraphics[width=0.25\textwidth]{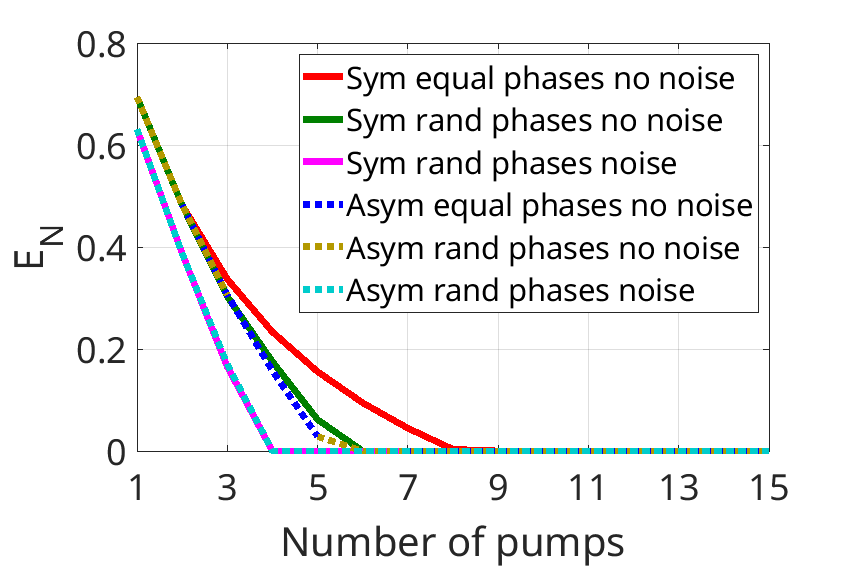}};
        \node at (-6.4, -0.6) {(a)};
        
        \node[anchor=north west] (img2) at (-3.25, 0) {\includegraphics[width=0.25\textwidth]{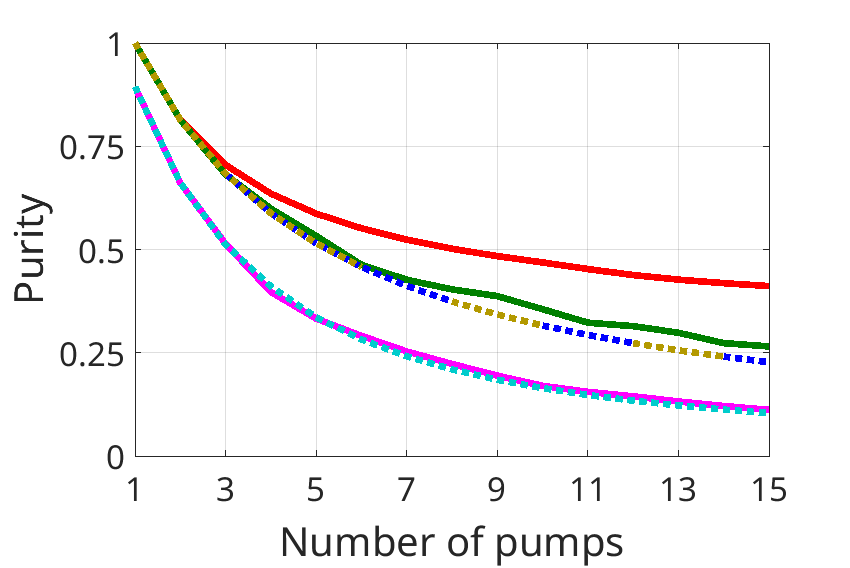}};
        \node at (-2.15, -0.6) {(b)};
    \end{tikzpicture}

\caption{Logarithmic negativity (a) and purity (b) of a 1, ..., 15 pump systems. We compare the theoretical predictions for symmetric (solid line, red, green, magenta) and asymmetric (dashed line, blue, yellow, cyan) systems with different parameter values. First, we have put all phases of the pumps equal to 0 and added no noise. Second, we have chosen all phases of the pumps randomly (random $\phi_k$ in the inset pump amplitude) and added no noise. Third, we have chosen all phases randomly and put a constant amount of noise $\theta=\frac{\pi}{4}$ via the quantum limited ($\bar{n}=1$) attenuator channel. All pumps have equal amplitudes at $\alpha=0.25\kappa$ in all cases. Note that the random phases have almost no effect on the asymmetric case, which can be seen from the almost perfectly overlaid blue and yellow plots. }
\label{Fig:constsum}
\end{figure}

We solved the conditional equations of motion for the second moments (see Eq. \ref{riccati}) under both symmetric and asymmetric parametric pumping. Unconditional dynamics failed to capture the experimental behavior, necessitating a conditional approach. By introducing loss via the attenuator channel in Eq. \ref{noisemodel}, the model reproduces the non-unit purity (for definition, see Ap. \ref{lognegappendix}) seen in experimental covariance matrices. In our theory, the squared magnitudes of the squeezing parameters $\alpha_k=|\alpha_k|\text{e}^{i\phi_k}$:s denote pump power. In the calculation, the pumps' phases, denoted by $\phi_k$, were chosen either at random or with a fixed phase relation were all were set to 0. The effects of all of the parameters on the theoretical prediction for the logarithmic negativity and purity can be seen from Fig. \ref{Fig:constsum}. There, we compare symmetric and asymmetric pumping configurations and the effects of varying the parameters. We solve the equations for both systems with all phases set to 0 and compare this with a system with all phases chosen completely randomly. We can see that in the symmetric case, different combinations of phases leads to destructive/constructive interference, resulting in reduced/improved $E_N$ or purity. Random phases most often lead to reduced negativity and purity. For the asymmetric system, phases make almost no difference for $E_N$ or purity.

Additionally, we applied a loss channel with $\theta = \pi/4$, corresponding to mixing the state on a 50/50 beam splitter with a thermal state, according to Eq. \ref{noisemodel}. This parameter has a degrading effect on $E_N$ as well as purity, as can be seen from Fig. \ref{Fig:constsum}. 

\subsection{Experiment}

Quadratures are measured by sampling the noise output from the flux pumped JPA, and covariance matrices are computed from the data. Gain calibrations $G_n$ for modes $n$ are obtained by the temperature stage and used to normalize the measured data according to formula: 

\begin{equation}
    \left(\boldsymbol{\sigma}_{on} \right)_{mn} = \frac{2 \langle (r_m - \langle r_m \rangle) (r_n - \langle r_n \rangle) \rangle}{\sqrt{G_mG_nf_mf_n}Z_0hB}, 
\end{equation}
where $r_n \in \{I_n, Q_n\}$, $f_n$ is measurement frequency, $Z_0=50\Omega$ the system impedance, $h$ Planck's constant and $B$ measurement bandwidth.

Noise contributions from sources other than the device itself can be subtracted by measuring the system noise with all pumps off. This also results in the quantum contribution being deducted, so it is added back according to formula:

\begin{equation}
    \boldsymbol{\sigma} = (\boldsymbol{\sigma}_{on} - \boldsymbol{\sigma}_{off})+\mathbf{I}\textrm{coth}\frac{hf_n}{2k_bT}, 
\end{equation}
where $k_b$ is Boltzmann's constant and T the physical temperature of the system, which in our experiment is below $T<20$mK, yielding $\textrm{coth}(...)\approx1.000$. The ground state is normalized to the identity matrix.

For the symmetric configuration, the pump spacing is $\Delta=0.1$ MHz (Fig. \ref{Fig:30 mode sym}), fully defining the mode structure after two pumps. Similarly, the asymmetric spacing is $\Delta_1=0.1$ MHz (Fig. \ref{Fig:asymmetric pumps}), and the rest are chosen randomly but of the same magnitude. In both cases, the two principal modes are measured at $\pm0.05$ MHz offsets from resonance. The pump amplitudes $|\alpha_E|$ used in the experiment are extracted from the onset of parametric oscillations in terms of coupling rate $\kappa$ (see App. \ref{calibrationappendix}). In comparison between theory and experiment, there is a mismatch between pump amplitudes, which we assign to the fact that not all degrees of freedom are monitored in the experiment. This leads to parametric oscillation instability that is not present in simulations. Consequently, the pump amplitude in the simulations is considered as a fit parameter in our comparisons.

Because the experimental results were obtained using random phases and the symmetric and asymmetric simulations show negligible differences, our subsequent comparisons focus mainly on the symmetric simulation.

\subsection{Comparison between theory and experiment}

\begin{figure}
    \centering
    \begin{tikzpicture}
        \node[anchor=north west] (img1) at (-7.5,0) {\includegraphics[width=0.25\textwidth]{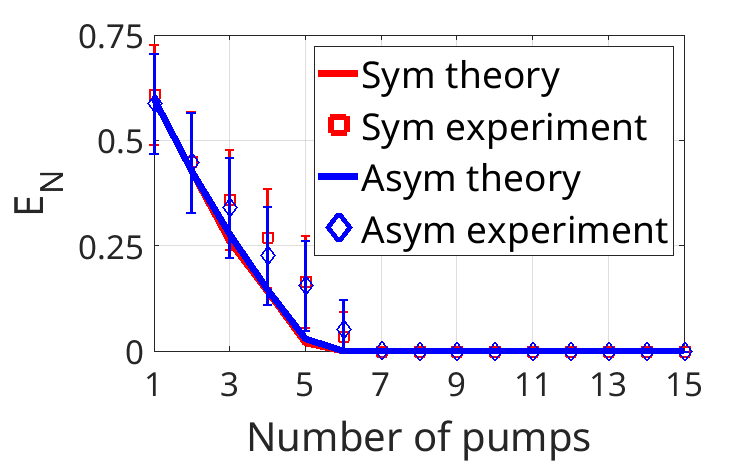}};
        \node at (-6.0, -0.6) {(a)};


        \node[anchor=north west] (img2) at (-3.25,-0.0) {\includegraphics[width=0.25\textwidth]{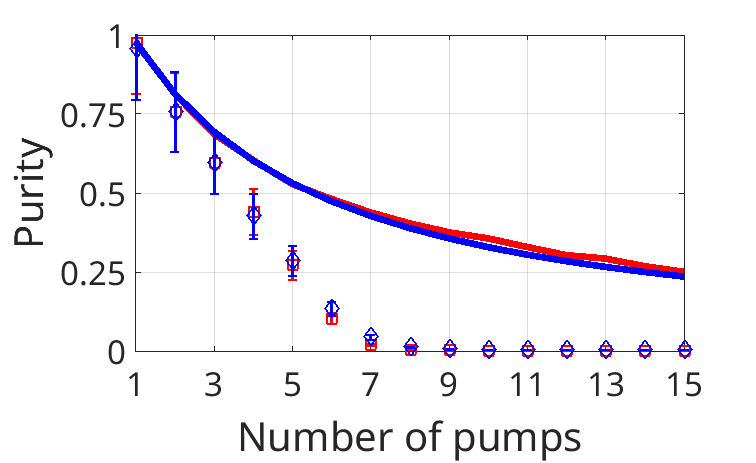}};
        \node at (-1.75, -0.6) {(b)};

    \end{tikzpicture}
\caption{Logarithmic negativity (a) and purity (b) of a 1, ..., 15 pump systems: solid curves for theory and symbols for the experimental results. First pump amplitude is fixed at $\alpha_1=0.22\kappa \ (\alpha_{E}=0.075\kappa)$ and the added pumps are $\alpha_j = 0.22\kappa \ (\alpha_{E}=0.075\kappa)$. The phases of all pumps are randomly chosen. The quantum limited ($\bar{n}=1$) attenuator channel with $\theta=\pi/8$ was applied. The strong decrease of purity in the experimental data is due to instability and the onset of oscillation as the number of pumps increases. The calculation with monitoring, however, does not display any instability.}
\label{fig:fixlow}
\end{figure}

\begin{figure}
    \centering
    \begin{tikzpicture}
        \node[anchor=north west] (img1) at (-7.5,0) {\includegraphics[width=0.25\textwidth]{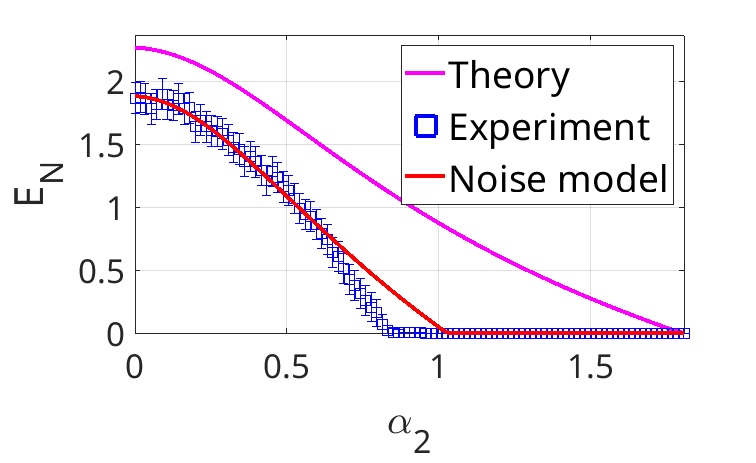}};
        \node at (-6, -0.6) {(a)};
        
        \node[anchor=north west] (img2) at (-3.25,-0.0) {\includegraphics[width=0.25\textwidth]{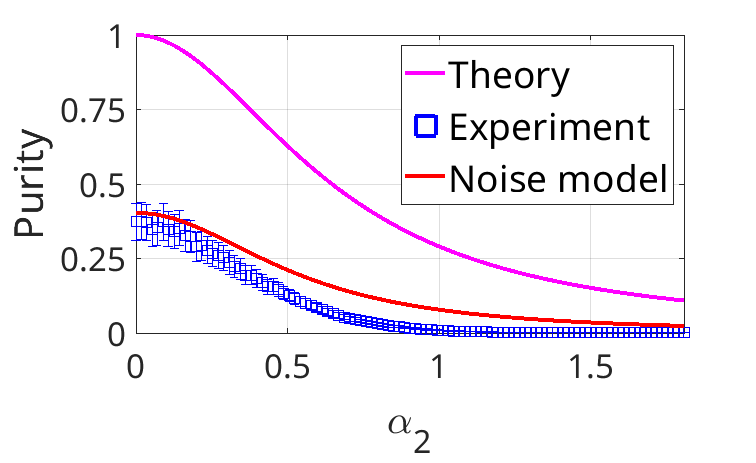}};
        \node at (-1.75, -0.6) {(b)};
    \end{tikzpicture}
\caption{Logarithmic negativity (a) and purity (b) of a 2 pump system where the first pump is fixed at $\alpha_1=1.15\kappa \ (\alpha_{E}=0.28\kappa)$ and the second pump grows from 0 to $1.81\kappa \ (\alpha_{E}=0 \dots 0.44\kappa)$. Phases of the pumps were chosen at random. One can notice that in the case with no noise, even when the second pump is at 0, experimental results show very low purity. To account for this, the quantum limited ($\bar{n}=1$) attenuator channel with $\theta = 2\pi/7$ was applied.}
\label{fig:amplitudesweep1}
\end{figure}

\begin{figure}
    \centering
    \begin{tikzpicture}
        \node[anchor=north west] (img1) at (-7.5,0) {\includegraphics[width=0.24\textwidth]{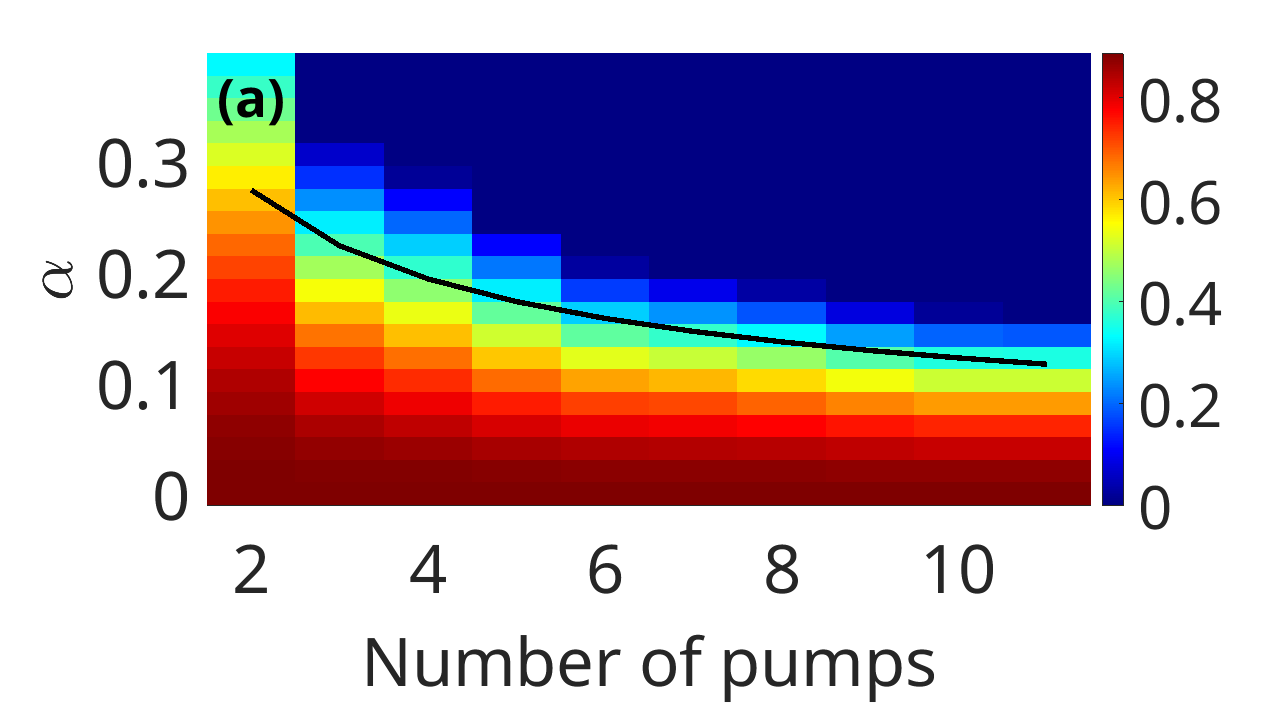}};
        
        \node[anchor=north west] (img2) at (-3.25,0) {\includegraphics[width=0.24\textwidth]{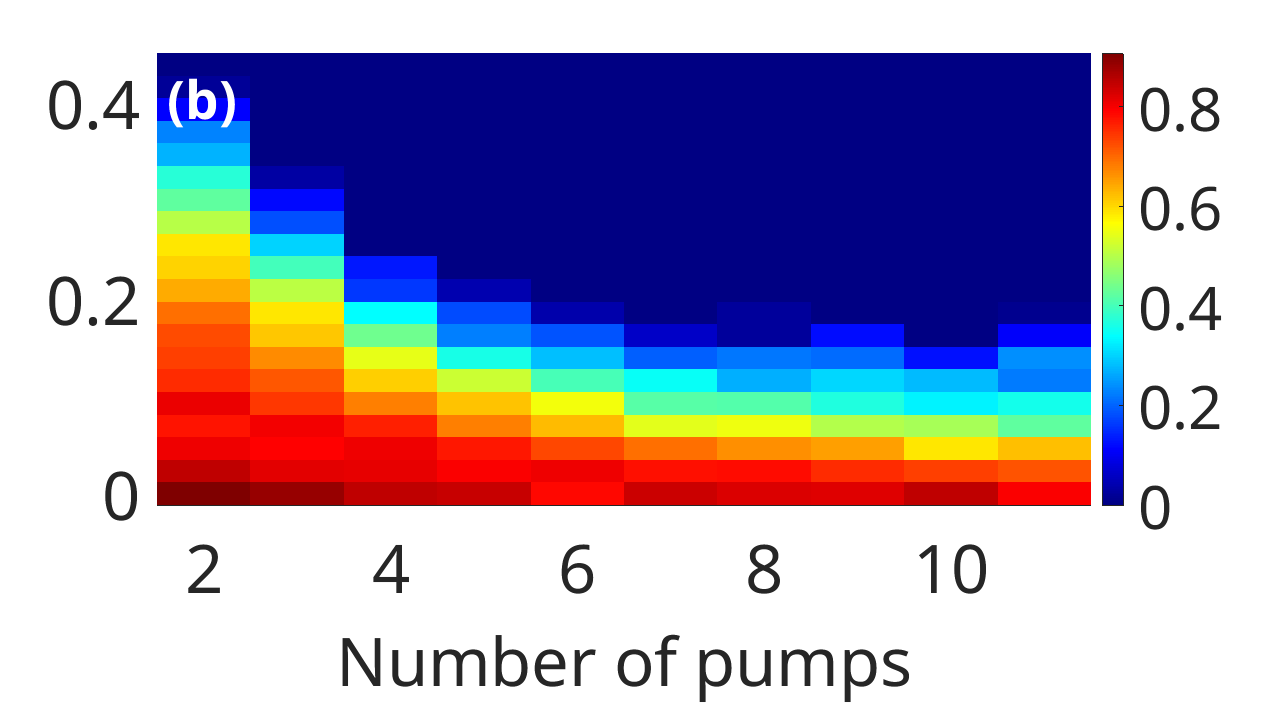}};
    \end{tikzpicture}
    \caption{Logarithmic negativity (scale given by the color bar) obtained from (a) theory and (b) experiment of a symmetrically pumped system with the first pump fixed at $\alpha_1 = 0.44\kappa \ (\alpha_{E}=0.15\kappa)$. In all configurations, pumps 2 through 11 share the same pump amplitude, which increases from $0$ to $0.44\kappa \ (\alpha_{E}=0 \dots 0.15\kappa)$. Phases of the pumps were chosen at random. Due to low purity in the experiment, we apply the quantum limited ($\bar{n}=1$) attenuator channel using $\theta = 2\pi/7$. The overlaid black curve in (a) is the logarithmic negativity of a system where each pump configuration has the same total power. }
    \label{fig:ampsweeplogneg}
\end{figure}

\begin{figure}
    \centering
    \begin{tikzpicture}
        \node[anchor=north west] (img1) at (-7.5,0) {\includegraphics[width=0.24\textwidth]{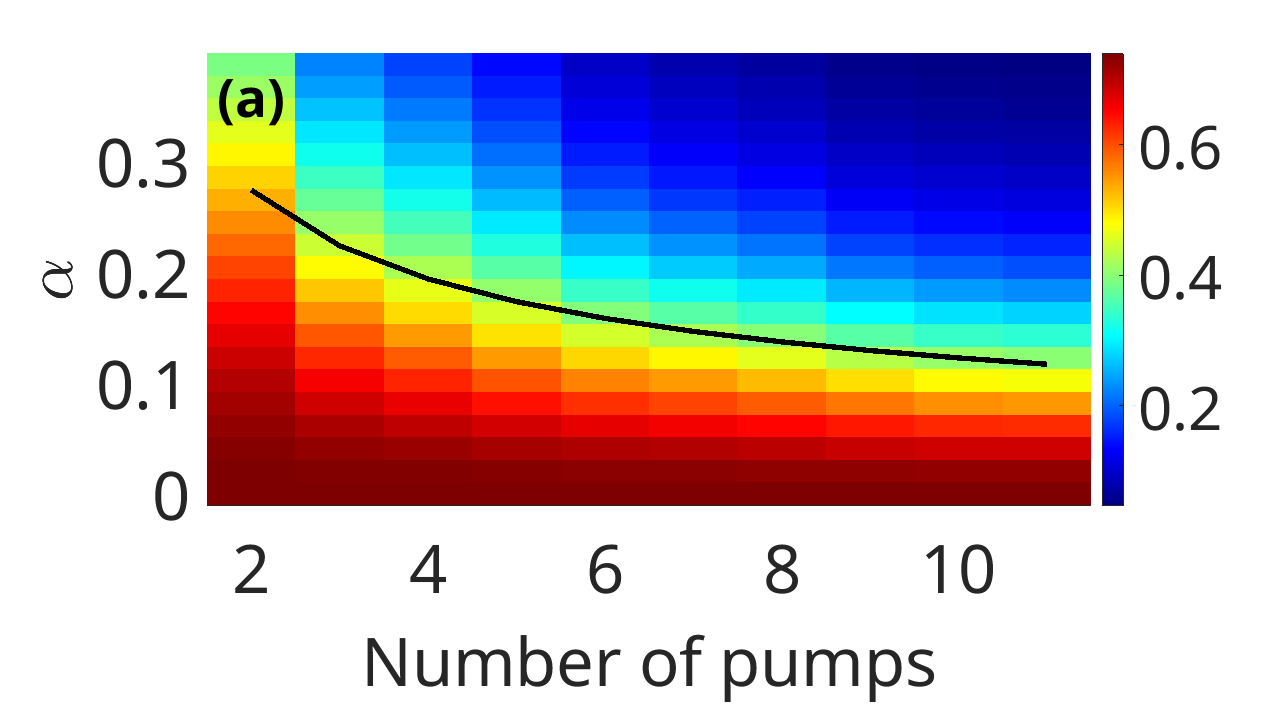}};
        
        \node[anchor=north west] (img2) at (-3.325,0) {\includegraphics[width=0.24\textwidth]{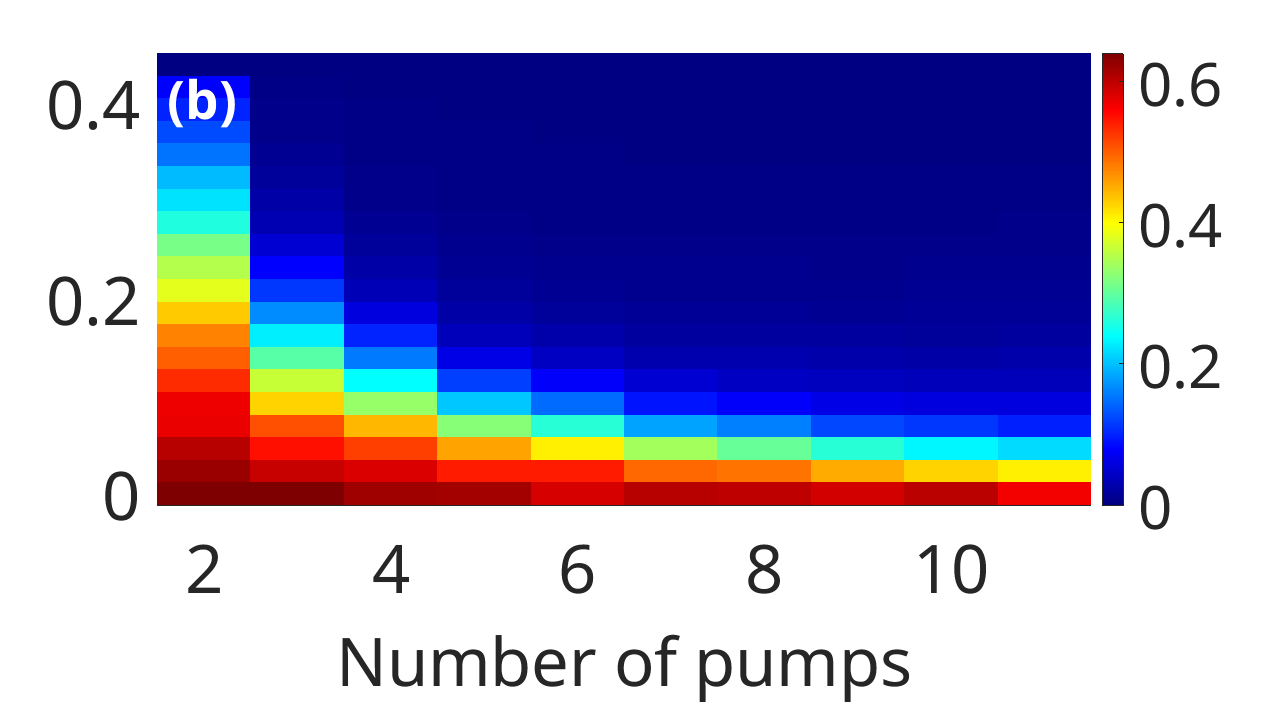}};
    \end{tikzpicture}
    \caption{Purity (scale given by the color bar) obtained from a) theory and b) experiment of symmetrically pumped system with first pump fixed at $\alpha_1 = 0.44\kappa \ (\alpha_{E}=0.15\kappa)$. In all configurations, all pumps 2,...,11 have the same pump amplitude that is grows from $0$ to $0.44\kappa \ (\alpha_{E}=0 \dots 0.15\kappa)$. Phases of the pumps were chosen at random. Due to low purity in the experiment, we apply the quantum limited ($\bar{n}=1$) attenuator channel with $\theta = 2\pi/7$. The overlaid black curve in (a) is the purity of a system where each pump configuration has the same total power.}
    \label{fig:purampsweep}
\end{figure}

The only parameters that we control in the experiment are the pump amplitudes $|\alpha_p|$. In the comparison between the experiment and theory in Figs. \ref{fig:fixlow} -- \ref{fig:purampsweep}, the first pump amplitude is adjusted to match the logarithmic negativity of the single pump case, because then the full state is completely separable into multiple TMS states, and the logarithmic negativity directly corresponds to the amount of squeezing. The rest are determined in the same manner as in the experiment.

The loss parameter $\theta$ is adjusted simultaneously with the amplitude calibration to match the purity of the single pump case. 
In the single pump case, tracing over unwanted modes does not decrease the purity, as the state is completely separable into multiple TMS states. Therefore, we must have some losses present in our system other than the growing number of idler modes. In addition, all phases were chosen randomly, as the relative phases of the pumps were not in our control in the experiment. In the simulations, no averaging over random phases was done.

Fig. \ref{fig:fixlow} displays logarithmic negativity and purity in a system of 1, ..., 15 pumps with random selection of phases. The first pump amplitude is fixed at $\alpha_1=0.22\kappa$, while the pumps $N=2 \dots 15$ have strengths $\alpha_j = 0.22\kappa$. Losses before the measurement and additional noise were modeled using attenuator channel with $\theta=\pi/8$ and $\bar{n}=1$. The strong decrease of purity in the experimental data is assigned to the onset of cavity oscillations due to parametric instability as the number of pumps increases. However, the calculation with heterodyne monitoring based on Eq. \ref{riccati} does not display any instability at $|\alpha| = \kappa/2$, even though it does appear in homodyne measurement and unmonitored cases. Monitoring stabilizes the dynamics and allows for good agreement for logarithmic negativity in Fig. \ref{fig:fixlow}a. The strong decrease in purity may be a sign that not all relevant modes are monitored in the experimental setting, and a parametric oscillation may start in the experiment, contrary to the expectation from the theory. At small number of pumps, i.e. at weak pump amplitude, the simulation for purity reproduces the measurement well.

To assess the influence of single pump strength versus several pumps, we performed amplitude sweeps with one pump  and multiple pumps (Figs. \ref{fig:amplitudesweep1} -- \ref{fig:purampsweep}). Fig. \ref{fig:amplitudesweep1} displays how $E_N$ for a two mode squeezed state induced by a pump with strength  $\alpha_1=1.15\kappa$ decreases when applying a second pump at offset frequency $\Delta{}_1 = +0.1$MHz. The amplitude of the second pump grows from 0 to $1.81\kappa$. Losses before the measurement and additional noise were modeled using attenuator channel with $\theta=2\pi/7$. The amount of losses can be assigned to the larger amount of power in this experiment.

Figs. \ref{fig:ampsweeplogneg} and \ref{fig:purampsweep} present theoretical (a) and experimental (b) amplitude sweep results for logarithmic negativity and purity for a TMS with initially $E_N=0.65$ and $\mu=0.78$, produced by a single fixed pump with $\alpha_1=0.44\kappa$. Additional pumps $N=2 \dots 11$ are applied and their amplitude is increased up to 0.44$\kappa$. We see no need to investigate the sweeping with larger number of pumps as the logarithmic negativity and purity drop very low already with smaller number of pumps, indicated by Figs. \ref{Fig:constsum} and \ref{fig:fixlow}. Losses before the measurement and additional noise were modeled using the attenuator channel with $\theta=2\pi/7$ due to the low initial purity. Our simulations suggest that the losses from the Gaussian channel in Eq. \ref{noisemodel} have a larger effect with larger power. This is why in Fig. \ref{fig:amplitudesweep1}b the purity is lower than in \ref{fig:purampsweep}. A clear decrease of $E_N$ and $\mu$ is observed with the increase of the number of pumps. The ovelaid black curves in theoretical plots \ref{fig:ampsweeplogneg}a and \ref{fig:purampsweep}a denote equipower contours given by the constant sum of squared pump amplitudes: $P = \sum_k |\alpha_k|^2/ N$, where $\alpha_k=\alpha$ is the amplitude of pump $k$ and $N$ is the number of pumps. The overlaid curves indicate that, even though the total applied power is constant, we still see the drop in $E_N$ and purity. 

\subsection{Discussion}

We chose the symmetric and asymmetric pumping configurations for two different properties: in the symmetric case, as correlations are created within a fixed mode set, we get an increasing number of BS correlations, and in the asymmetric case we get rid of those, but are left with a huge number of idler modes as a trade-off.
Even though we chose these two specific configurations, the analysis can in principle be applied to any pumping topology. 

In the calculation, the pumps' phases were chosen either at random or with all phases set to 0. The phases determine the relative signs of the squeezing parameters, resulting in constructive/destructive interference, which has an observable effect for the symmetric case. Our calculations suggest that this effect is as strong as the number of idlers in the reduction of bipartite entanglement, as can be seen from Fig. \ref{Fig:constsum}. Using the phases set to zero for all pump tones, there was a clear difference between symmetric and asymmetric pump configurations: both the logarithmic negativity $E_N$ and purity $\mu$ decayed slower with the increasing number of pumps. The simulation using random phases degraded the results of symmetric pump distribution to a level that was hardly distinguishable from the asymmetric case. The results for the random phase selection are presented without averaging over different phase configurations, leading to a clear scatter between adjacent points (see Fig. \ref{Fig:constsum}). Investigating the phase dependence further is outside the scope of our work. 

For identical phases and no noise, symmetric and asymmetric systems differ markedly when there are more than 2 pumps. In the symmetric system with 1 or 2 pumps, random phases make almost no difference. In the asymmetric case, we are effectively making new copies of the 2-pump system that do not couple to one another directly due to the simplifying assumptions. This explains why random phases have no degrading effect on the asymmetric case. 

In the symmetric configuration, pump tones couple only among the 44 system modes. As more pumps are added, bipartite entanglement between a reference mode pair decreases, despite the network size remaining fixed (see Sect. \ref{GraphTheory}). This results from increased TMS correlations that redistribute entanglement across the network. Our simulations based on Eq. \ref{riccati} and the experimental results confirm this trend.

In the asymmetric case (see Fig. \ref{Fig:constsum}), each additional pump with random phase connects system modes to new extra modes, which are assumed to be non-interacting. Despite the growing Hilbert space, entanglement between the reference pair (–1, 1) decays nearly identically with the symmetric case, indicating the loss of bipartite entanglement rather by correlation redistribution than mode count. This is evidenced by increasing diagonal elements and decreasing off-diagonal elements in sub-covariance matrices.

Increasing pump power leads to small pump-induced resonance shifts and weak mode hybridization. Pump induced mode hybridization is not relevant for our case, because a circulator assures that no energy is reflected back and is flooded out. These effects could possibly be incorporated through effective detuning parameters extracted from the measured spectra. Their inclusion might improve quantitative agreement but does not alter the qualitative trend of entanglement redistribution. This kind of parameter fitting is outside the scope of this work. Higher pumping power may lead to excitations in two-level systems inside the cavity, but these effects are expected to primarily introduce additional loss and dephasing channels rather than fundamentally modify the underlying entanglement redistribution mechanism. 

Overall, our analysis has identified three key mechanisms behind the observed reduction of bipartite entanglement with increasing number of pumps: (i) reduced effective two-mode squeezing due to a decrease of the off-diagonal elements because of entanglement redistribution and (ii) increasing number of edge-correlations in the graphs leading to larger loss of purity while tracing over unwanted modes, leading to an increase of the diagonal elements, and (iii) pre-measurement losses arising from the finite transmission efficiecy between the device and the measurement, including added noise from the amplification chain. 

\section{Conclusions} \label{Conclusion}
Our investigation of multiple parametric pump tones in two-mode squeezing (TMS) revealed fundamental difficulties in the behavior of bipartite entanglement in Josephson parametric circuits. Through the solution of the conditional equations of motion for second moments in both symmetric and asymmetric pump configurations, we showed that the addition of pumps does not only introduce new TMS states but also generates correlations among all modes, redistributing bipartite correlations, thus leading to a reduction in the entanglement of individual mode pairs. Overall, our experimental results match theoretical predictions, with discrepancies primarily attributed to internal losses in the Josephson parametric amplifier, which are not included in the model we used in this paper. Our findings indicate that maintaining high entanglement between specific mode pairs becomes increasingly challenging as more pumps are introduced, as quantum information is redistributed across a larger network of modes.  
\vspace{20pt}

\section*{Acknowledgments}
The authors thank Visa Vesterinen and VTT Technical Research Center of Finland for generous support with the Josephson traveling wave parametric amplifier and to Esko Keski-Vakkuri, Christian Flindt, Kimmo Luoma, and Alexander Zyuzin for fruitful discussions. The experimental work benefited from the Aalto University OtaNano/LTL infrastructure. 

\section*{Funding}

This work was supported by the Academy of Finland (AF) projects 341913 (EFT), and 312295 \& 352926 (CoE, Quantum Technology Finland). The research leading to these results has received funding from the European Union’s Horizon 2020 Research and Innovation Programme, under Grant Agreement No.~824109 (EMP). The work of KP was supported by the European Union under Horizon Europe 2021-2027 Framework Programme, project MiSS (Microwave Squeezing with Superconducting (meta)materials) grant agreement ID: 101135868. IL is grateful for Vaisala Foundation of the Finnish Academy of Arts and Letters for a stipend. EM acknowledges a PhD scholarship from InstituteQ. The support of the Jane and Aatos Erkko Foundation and the Keele Foundation (SuperC project) is also gratefully acknowledged.



\section*{Competing interests}

The authors declare no competing interests for this work. 

\appendix
\counterwithin{figure}{section}

\section{Placement of modes and pump frequencies within the JPA resonance}
\label{mode_posit}
\begin{figure}[htb!]
    \centering
    \includegraphics[width=0.85\linewidth]{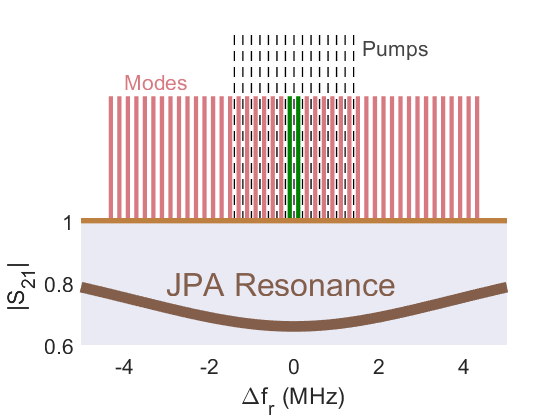}
    \caption{Placement of modes and pumps. The first modes, depicted in as green shaded regions, (-1,1) are set at [-0.05, 0.05] MHz detuning from the resonance frequency (see Fig. \ref{Fig:30 mode sym} for the indexing of the modes). This spacing means a full span of $2\times(0.05 + 21 \times 0.1)=4.3$ MHz. Red shaded areas depict mode locations taken into account in the simulation, while dashed black lines depict pump placement.)}
    \label{fig:resonance}
\end{figure}

\section{The adjacency matrix}
\label{adjacency}
The adjacency matrix \label{adjacency matrix} is used to indicate which pairs of nodes are \textit{adjacent} in a graph, that is, if they are connected by an edge or not. It is a square matrix whose indices represent nodes and nonzero elements represent edges with weights. A formal definition \cite{handbook}: Let $U=\{u_1, u_2, ..., u_n\}$ be a set of nodes, $E=\{(u_i, u_j), ...\}$ a set of edges, and $\alpha=\{\alpha_{ij}, ...\}$ a set of weights. An element $(\text{Adj} \mathbf{M})_{ij}$ of the adjacency matrix is $\alpha_{ij}$, if there is an edge of weight $\alpha_{ij}$ between the nodes $u_i$ and $u_j$: 
\begin{equation}
    (\text{Adj} \mathbf{M})_{ij} = 
\begin{cases}
& \alpha_{ij} \quad \text{if} \; \; (u_i, u_j) \; \in \; E\\
& 0 \quad \text{otherwise}
\end{cases}
\end{equation}

\section{Logarithmic negativity and purity}
\label{lognegappendix}
We are interested in the entanglement between the pair of modes (-1, 1). Therefore, we took the submatrix of these two modes from the full covariance matrix and analyzed its properties. The entanglement monotone we used was the logarithmic negativity, which is both a necessary and sufficient criterion for 2-mode Gaussian systems. It is obtained from \cite{logneg, volofesep}:

\begin{equation}
    E_N = \max\{0, -\log_2  \tilde{\nu}_{-}\},
\end{equation}
where $\tilde{\nu}_{-}$ is the smallest partially transposed eigenvalue 

\begin{equation}
    \tilde{\nu}_{\pm} = \frac{\tilde{\Delta}\pm\sqrt{\tilde{\Delta}^2-4 \text{Det}\boldsymbol{\sigma}}}{2},
\end{equation}
using the following definitions:
\begin{equation}
    \tilde{\Delta} = \text{Det}\boldsymbol{\sigma}_{-1} + \text{Det}\boldsymbol{\sigma}_{1} - 2\text{Det}\boldsymbol{\sigma}_{-1, 1}, \hspace{5pt}
    \boldsymbol{\sigma}=\begin{pmatrix}
    \boldsymbol{\sigma}_{-1}&\boldsymbol{\sigma}_{-1, 1}\\
    \boldsymbol{\sigma}_{-1, 1}^{\text{T}}&\boldsymbol{\sigma}_{1}
    \end{pmatrix}.
\end{equation}

Purity is defined by: 
\begin{equation}
    \mu = \frac{1}{\sqrt{\text{Det}\mathbf{\sigma}}}.
\end{equation}
\section{Calibration of the critical point in experiment}
\label{calibrationappendix}
\begin{figure}[ht!]
    \centering
    \includegraphics[width=0.85\linewidth]{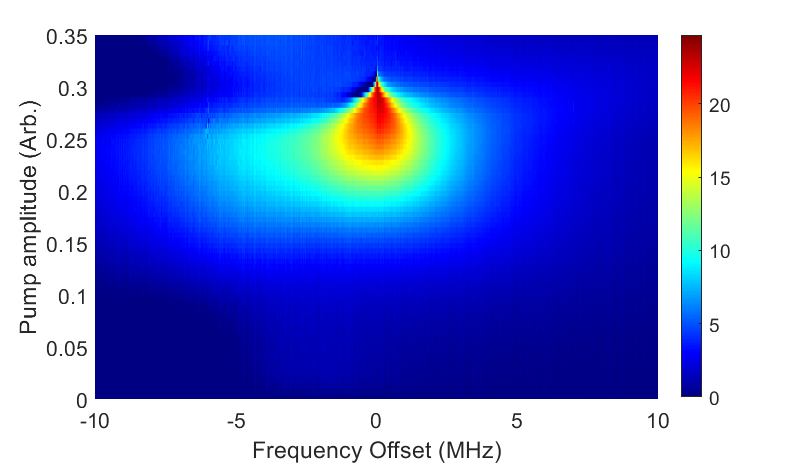}
    \caption{Gain of the JPA as a function of the pump amplitude and the frequency offset from half of the pump frequency $\omega_{\rm{p}} /2$. Critical pump amplitude is taken from the top of the drop-shaped gain region. Scale bar for gain on the right is given in dB. 
    The system starts to oscillate at the critical amplitude $\alpha=0.315-0.32 \rm{\;(a.u.)}$ which is taken to correspond to $\alpha=0.5 \kappa$. }
    \label{fig:calibration}
\end{figure}

\bibliography{Sources,references,apssamp}

\end{document}